\documentclass[twocolumn,letterpaper,aps,prc,superscriptaddress,nofootinbib,showpacs,floatfix,longbibliography]{revtex4-1}

\usepackage{graphicx}	
\usepackage{xspace}     
\usepackage{amsmath}

\newcommand{\pt}{\mbox{$p_T$}\xspace}
\newcommand{\pte}{\mbox{${p_T}^{e}$}\xspace}
\newcommand{\raa}{\mbox{$R_{AA}$}\xspace}
\newcommand{\rcp}{\mbox{$R_{CP}$}\xspace}

\newcommand{\Ncoll}{\mbox{$N_{\rm coll}$}\xspace}

\newcommand{\sqsn}{\mbox{$\sqrt{s_{_{NN}}}$}\xspace}
\newcommand{\sqsnsix}{\mbox{$\sqrt{s_{_{NN}}}=62.4$~GeV}\xspace}
\newcommand{\sqsntwo}{\mbox{$\sqrt{s_{_{NN}}}=200$~GeV}\xspace}
\newcommand{\vtwo}{\mbox{$v_2$}\xspace}
\newcommand{\pp}{\mbox{$p$$+$$p$}\xspace}
\newcommand{\auau}{\mbox{Au$+$Au}\xspace}
\newcommand{\pizero}{\mbox{$\pi^0$}\xspace}
\newcommand{\vtworaw}{\mbox{${{v_2}^{{\rm raw}}}$}\xspace}

\newcommand{\rsb}{\mbox{$S/B$}\xspace}
\newcommand{\dAu}{\mbox{$d$$+$Au}\xspace}
\newcommand{\dA}{\mbox{$d$$+$$A$}\xspace}

\begin{document}

\title{Heavy-quark production and elliptic flow in Au$+$Au collisions at 
$\sqrt{s_{_{NN}}}=62.4$~GeV}

\newcommand{\abilene}{Abilene Christian University, Abilene, Texas 79699, USA}
\newcommand{\augie}{Department of Physics, Augustana College, Sioux Falls, South Dakota 57197, USA}
\newcommand{\banaras}{Department of Physics, Banaras Hindu University, Varanasi 221005, India}
\newcommand{\barc}{Bhabha Atomic Research Centre, Bombay 400 085, India}
\newcommand{\baruch}{Baruch College, City University of New York, New York, New York, 10010 USA}
\newcommand{\bnlcoll}{Collider-Accelerator Department, Brookhaven National Laboratory, Upton, New York 11973-5000, USA}
\newcommand{\bnlphys}{Physics Department, Brookhaven National Laboratory, Upton, New York 11973-5000, USA}
\newcommand{\caucr}{University of California - Riverside, Riverside, California 92521, USA}
\newcommand{\charlesczech}{Charles University, Ovocn\'{y} trh 5, Praha 1, 116 36, Prague, Czech Republic}
\newcommand{\chonbuk}{Chonbuk National University, Jeonju, 561-756, Korea}
\newcommand{\cns}{Center for Nuclear Study, Graduate School of Science, University of Tokyo, 7-3-1 Hongo, Bunkyo, Tokyo 113-0033, Japan}
\newcommand{\colorado}{University of Colorado, Boulder, Colorado 80309, USA}
\newcommand{\columbia}{Columbia University, New York, New York 10027 and Nevis Laboratories, Irvington, New York 10533, USA}
\newcommand{\czechtech}{Czech Technical University, Zikova 4, 166 36 Prague 6, Czech Republic}
\newcommand{\dapnia}{Dapnia, CEA Saclay, F-91191, Gif-sur-Yvette, France}
\newcommand{\debrecen}{Debrecen University, H-4010 Debrecen, Egyetem t{\'e}r 1, Hungary}
\newcommand{\elte}{ELTE, E{\"o}tv{\"o}s Lor{\'a}nd University, H - 1117 Budapest, P{\'a}zm{\'a}ny P. s. 1/A, Hungary}
\newcommand{\ewha}{Ewha Womans University, Seoul 120-750, Korea}
\newcommand{\fsu}{Florida State University, Tallahassee, Florida 32306, USA}
\newcommand{\gsu}{Georgia State University, Atlanta, Georgia 30303, USA}
\newcommand{\hanyang}{Hanyang University, Seoul 133-792, Korea}
\newcommand{\hiroshima}{Hiroshima University, Kagamiyama, Higashi-Hiroshima 739-8526, Japan}
\newcommand{\ihepprot}{IHEP Protvino, State Research Center of Russian Federation, Institute for High Energy Physics, Protvino, 142281, Russia}
\newcommand{\illuiuc}{University of Illinois at Urbana-Champaign, Urbana, Illinois 61801, USA}
\newcommand{\inrras}{Institute for Nuclear Research of the Russian Academy of Sciences, prospekt 60-letiya Oktyabrya 7a, Moscow 117312, Russia}
\newcommand{\instpasczech}{Institute of Physics, Academy of Sciences of the Czech Republic, Na Slovance 2, 182 21 Prague 8, Czech Republic}
\newcommand{\isu}{Iowa State University, Ames, Iowa 50011, USA}
\newcommand{\jyvaskyla}{Helsinki Institute of Physics and University of Jyv{\"a}skyl{\"a}, P.O.Box 35, FI-40014 Jyv{\"a}skyl{\"a}, Finland}
\newcommand{\kek}{KEK, High Energy Accelerator Research Organization, Tsukuba, Ibaraki 305-0801, Japan}
\newcommand{\korea}{Korea University, Seoul, 136-701, Korea}
\newcommand{\kurchatov}{Russian Research Center ``Kurchatov Institute", Moscow, 123098 Russia}
\newcommand{\kyoto}{Kyoto University, Kyoto 606-8502, Japan}
\newcommand{\labllr}{Laboratoire Leprince-Ringuet, Ecole Polytechnique, CNRS-IN2P3, Route de Saclay, F-91128, Palaiseau, France}
\newcommand{\lahorelums}{Physics Department, Lahore University of Management Sciences, Lahore, Pakistan}
\newcommand{\lawllnl}{Lawrence Livermore National Laboratory, Livermore, California 94550, USA}
\newcommand{\losalamos}{Los Alamos National Laboratory, Los Alamos, New Mexico 87545, USA}
\newcommand{\lpc}{LPC, Universit{\'e} Blaise Pascal, CNRS-IN2P3, Clermont-Fd, 63177 Aubiere Cedex, France}
\newcommand{\lund}{Department of Physics, Lund University, Box 118, SE-221 00 Lund, Sweden}
\newcommand{\maryland}{University of Maryland, College Park, Maryland 20742, USA}
\newcommand{\mass}{Department of Physics, University of Massachusetts, Amherst, Massachusetts 01003-9337, USA }
\newcommand{\michigan}{Department of Physics, University of Michigan, Ann Arbor, Michigan 48109-1040, USA}
\newcommand{\muhlenberg}{Muhlenberg College, Allentown, Pennsylvania 18104-5586, USA}
\newcommand{\myongji}{Myongji University, Yongin, Kyonggido 449-728, Korea}
\newcommand{\nagasaki}{Nagasaki Institute of Applied Science, Nagasaki-shi, Nagasaki 851-0193, Japan}
\newcommand{\natmephi}{National Research Nuclear University, MEPhI, Moscow Engineering Physics Institute, Moscow, 115409, Russia}
\newcommand{\newmex}{University of New Mexico, Albuquerque, New Mexico 87131, USA }
\newcommand{\nmsu}{New Mexico State University, Las Cruces, New Mexico 88003, USA}
\newcommand{\ohio}{Department of Physics and Astronomy, Ohio University, Athens, Ohio 45701, USA}
\newcommand{\ornl}{Oak Ridge National Laboratory, Oak Ridge, Tennessee 37831, USA}
\newcommand{\orsay}{IPN-Orsay, Universite Paris Sud, CNRS-IN2P3, BP1, F-91406, Orsay, France}
\newcommand{\pnpi}{PNPI, Petersburg Nuclear Physics Institute, Gatchina, Leningrad region, 188300, Russia}
\newcommand{\riken}{RIKEN Nishina Center for Accelerator-Based Science, Wako, Saitama 351-0198, Japan}
\newcommand{\rikjrbrc}{RIKEN BNL Research Center, Brookhaven National Laboratory, Upton, New York 11973-5000, USA}
\newcommand{\rikkyo}{Physics Department, Rikkyo University, 3-34-1 Nishi-Ikebukuro, Toshima, Tokyo 171-8501, Japan}
\newcommand{\saispbstu}{Saint Petersburg State Polytechnic University, St. Petersburg, 195251 Russia}
\newcommand{\saopaulo}{Universidade de S{\~a}o Paulo, Instituto de F\'{\i}sica, Caixa Postal 66318, S{\~a}o Paulo CEP05315-970, Brazil}
\newcommand{\seoulnat}{Department of Physics and Astronomy, Seoul National University, Seoul, Korea}
\newcommand{\stonybrkc}{Chemistry Department, Stony Brook University, SUNY, Stony Brook, New York 11794-3400, USA}
\newcommand{\stonycrkp}{Department of Physics and Astronomy, Stony Brook University, SUNY, Stony Brook, New York 11794-3800,, USA}
\newcommand{\tenn}{University of Tennessee, Knoxville, Tennessee 37996, USA}
\newcommand{\titech}{Department of Physics, Tokyo Institute of Technology, Oh-okayama, Meguro, Tokyo 152-8551, Japan}
\newcommand{\tsukuba}{Institute of Physics, University of Tsukuba, Tsukuba, Ibaraki 305, Japan}
\newcommand{\vandy}{Vanderbilt University, Nashville, Tennessee 37235, USA}
\newcommand{\weizmann}{Weizmann Institute, Rehovot 76100, Israel}
\newcommand{\wigner}{Institute for Particle and Nuclear Physics, Wigner Research Centre for Physics, Hungarian Academy of Sciences (Wigner RCP, RMKI) H-1525 Budapest 114, POBox 49, Budapest, Hungary}
\newcommand{\yonsei}{Yonsei University, IPAP, Seoul 120-749, Korea}
\newcommand{\zagreb}{University of Zagreb, Faculty of Science, Department of Physics, Bijeni\v{c}ka 32, HR-10002 Zagreb, Croatia}
\affiliation{\abilene}
\affiliation{\augie}
\affiliation{\banaras}
\affiliation{\barc}
\affiliation{\baruch}
\affiliation{\bnlcoll}
\affiliation{\bnlphys}
\affiliation{\caucr}
\affiliation{\charlesczech}
\affiliation{\chonbuk}
\affiliation{\cns}
\affiliation{\colorado}
\affiliation{\columbia}
\affiliation{\czechtech}
\affiliation{\dapnia}
\affiliation{\debrecen}
\affiliation{\elte}
\affiliation{\ewha}
\affiliation{\fsu}
\affiliation{\gsu}
\affiliation{\hanyang}
\affiliation{\hiroshima}
\affiliation{\ihepprot}
\affiliation{\illuiuc}
\affiliation{\inrras}
\affiliation{\instpasczech}
\affiliation{\isu}
\affiliation{\jyvaskyla}
\affiliation{\kek}
\affiliation{\korea}
\affiliation{\kurchatov}
\affiliation{\kyoto}
\affiliation{\labllr}
\affiliation{\lahorelums}
\affiliation{\lawllnl}
\affiliation{\losalamos}
\affiliation{\lpc}
\affiliation{\lund}
\affiliation{\maryland}
\affiliation{\mass}
\affiliation{\michigan}
\affiliation{\muhlenberg}
\affiliation{\myongji}
\affiliation{\nagasaki}
\affiliation{\natmephi}
\affiliation{\newmex}
\affiliation{\nmsu}
\affiliation{\ohio}
\affiliation{\ornl}
\affiliation{\orsay}
\affiliation{\pnpi}
\affiliation{\riken}
\affiliation{\rikjrbrc}
\affiliation{\rikkyo}
\affiliation{\saispbstu}
\affiliation{\saopaulo}
\affiliation{\seoulnat}
\affiliation{\stonybrkc}
\affiliation{\stonycrkp}
\affiliation{\tenn}
\affiliation{\titech}
\affiliation{\tsukuba}
\affiliation{\vandy}
\affiliation{\weizmann}
\affiliation{\wigner}
\affiliation{\yonsei}
\affiliation{\zagreb}
\author{A.~Adare} \affiliation{\colorado}
\author{C.~Aidala} \affiliation{\losalamos} \affiliation{\michigan}
\author{N.N.~Ajitanand} \affiliation{\stonybrkc}
\author{Y.~Akiba} \affiliation{\riken} \affiliation{\rikjrbrc}
\author{R.~Akimoto} \affiliation{\cns}
\author{H.~Al-Ta'ani} \affiliation{\nmsu}
\author{J.~Alexander} \affiliation{\stonybrkc}
\author{A.~Angerami} \affiliation{\columbia}
\author{K.~Aoki} \affiliation{\riken}
\author{N.~Apadula} \affiliation{\stonycrkp}
% ...
\collaboration{PHENIX Collaboration} \noaffiliation

\date{\today}


\begin{abstract}


We present measurements of electrons and positrons from the semileptonic 
decays of heavy-flavor hadrons at midrapidity ($|y|<$ 0.35) in Au$+$Au 
collisions at $\sqrt{s_{_{NN}}}=62.4$~GeV. The data were collected in 2010 
by the PHENIX experiment that included the new hadron-blind detector. The 
invariant yield of electrons from heavy-flavor decays is measured as a 
function of transverse momentum in the range $1<p_T^e<5$~GeV/$c$.  
The invariant yield per binary collision is slightly enhanced above the 
$p$$+$$p$ reference in Au$+$Au 0\%--20\%, 20\%--40\% and 40\%--60\% 
centralities at a comparable level. At this low beam energy this may be a result of the interplay 
between initial-state Cronin effects, final-state flow, and energy loss 
in medium. The $v_2$ of electrons 
from heavy-flavor decays is nonzero when averaged between 
$1.3<p_T^e<2.5$~GeV/$c$ for 0\%--40\% centrality collisions at 
$\sqrt{s_{_{NN}}}=62.4$~GeV. For 20\%--40\% centrality collisions, 
the $v_2$ at $\sqrt{s_{_{NN}}}=62.4$~GeV is smaller than that for heavy 
flavor decays at $\sqrt{s_{_{NN}}}=200$~GeV.  The $v_2$ of the electrons 
from heavy-flavor decay at the lower beam energy is also smaller than 
$v_2$ for pions.  Both results indicate that the heavy-quarks interact 
with the medium formed in these collisions, but they may not be at the 
same level of thermalization with the medium as observed at 
$\sqrt{s_{_{NN}}}=200$~GeV.

\end{abstract}


	
\maketitle

\section{Introduction \label{sec01}}

Collisions of large nuclei at ultra-relativistic energies produce a state 
of matter, known as the quark-gluon plasma (QGP), in which the quarks and 
gluons that are normally bound inside hadrons become deconfined.  At the 
Relativistic Heavy Ion Collider (RHIC), collisions of heavy nuclei at 
\sqsntwo produce strongly coupled, dense partonic matter that exhibits 
strong collective motion~\cite{Adcox:2004mh}.  Comparisons of the measured 
anisotropic flow parameter \vtwo with hydrodynamic calculations indicate 
that the medium expands and flows as a near-perfect 
liquid~\cite{Huovinen:2001cy,Adler:2003kt,Adams:2003am}. 
The significant suppression of high-\pt particles produced in these collisions 
relative to scaled \pp collisions at the same center of mass energy also 
implies that partons lose energy while traversing the 
medium~\cite{Adcox:2001jp,Adler:2003qi,Adams:2003kv}. Both results 
indicate the formation of the QGP at \sqsntwo. It is important to map out 
these two key observations as a function of collision energy to study the 
transition from normal hadronic matter to the QGP.

Due to the short lifetime of the hot nuclear medium ($\sim10$~fm/$c$), 
experimental probes of the medium properties must be self-generated during 
the collision. To explore the formation and properties of strongly interacting
matter at lower energy density, a particularly useful set of probes is charm and bottom 
quarks.  At RHIC energies these quarks are produced primarily through 
gluon fusion in the initial stage of the collision, and are therefore 
present for the full evolution of the system, in contrast to the lighter 
quarks that can be produced thermally throughout the lifetime of the 
medium. Prior experiments have established that electrons from heavy 
flavor meson decays display a significant \vtwo in Au$+$Au collisions at 
\sqsntwo and Pb$+$Pb collisions at \sqsn$=2.76$~TeV, indicating that heavy 
quarks may experience collective motion along with the lighter partons 
that constitute the bulk of the 
medium~\cite{Adare:2006nq,Adare:2010de,Abelev:2013lca}.  In contrast with 
early predictions~\cite{Dokshitzer:2001zm,Armesto:2005iq}, heavy flavor 
hadrons are also significantly suppressed in central Au$+$Au collisions at 
\sqsntwo, at a level comparable to light-flavor 
hadrons~\cite{Adare:2006nq,Adare:2010de}. The magnitude of the suppression 
and flow of heavy quarks have proven to be a challenge to many models of 
parton energy loss in 
QGP~\cite{Sharma:2009hn,PhysRevC.86.014903,Uphoff:2012gb,Gossiaux:2008jv}.

To explore the formation and properties of lower energy density strongly 
interacting matter, Au$+$Au collisions with lower center of mass energies 
(62.4, 39, 11.5. and 7.7~GeV) were recorded during the 2010 RHIC run.  It 
was observed that inclusive hadrons and identified light-flavor hadrons 
display significant flow in Au$+$Au collisions at 
\sqsnsix~\cite{PhysRevLett.110.142301,ppg124}. 
However, the observed $\pi^{0}$ suppression 
is smaller than in higher energy collisions~\cite{PhysRevLett.109.152301} 
for $\pt<6$~GeV/$c$.  This may be due to a change in the competition 
between the Cronin enhancement that is prevalent in lower energy 
collisions and the suppressing effects of the hot 
medium~\cite{Vitev:2004gn}.  Cronin enhancement is also observed for 
electrons from heavy-flavor decays in $d$$+$Au collisions at 
$\sqrt{s_{_{NN}}}=200$~GeV~\cite{PhysRevLett.109.242301}, and is expected 
to be larger at lower energies~\cite{PhysRevLett.68.452}.

To provide more information on the formation and properties of the plasma 
produced at \sqsnsix at RHIC, and the possible role of initial-state 
effects, this paper presents measurements of the \pt spectra and 
anisotropic flow parameter \vtwo of electrons from the decays of heavy 
flavor (charm and beauty) hadrons produced in Au$+$Au collisions.

\section{Experiment Setup}
\label{sec02_setup}

\begin{figure}[thb]
\includegraphics[width=1.0\linewidth]{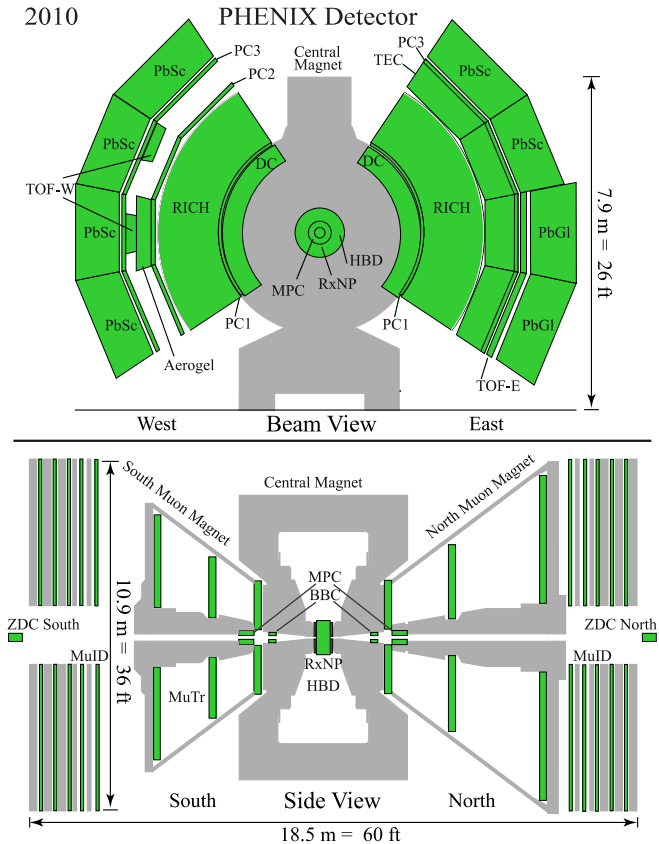}  
\caption{\label{fig:phenix_2010}(Color online) 
The PHENIX detector configuration for the 2010 data taking period. 
The upper panel is the beam view and the lower panel is the side view.
}
\end{figure}

PHENIX collected approximately 400 million events in 2010 for \auau 
collisions at \sqsnsix within $\pm20$~cm of the nominal collision point. 
Figure~\ref{fig:phenix_2010} shows the PHENIX detector system during the 
2010 data taking period. Details about PHENIX detector subsystems can be 
found in 
Refs.~\cite{Adcox2003469,Allen2003549,Richardson:2010hm,Fraenkel:2005wx,Anderson:2011jw,Adcox2003489,Aizawa2003508,Akiba:1999rs,Aphecetche2003521}.


The beam-beam counters (BBC) provide the measurement of collision time, 
collision vertex position along the beam axis, and the minimum-bias (MB) 
trigger information~\cite{Allen2003549}. BBCs are installed along the beam 
axis 144 cm from the center of PHENIX with a rapidity coverage of $3.0 < 
|\eta| < 3.9$.  The difference of the average hit time of PMTs between the 
North and the South BBC determines the collision vertex position along the 
beam direction, producing a vertex resolution in the beam direction of 
$\sim$0.5 cm in central \auau collisions.  The event centrality is also 
determined by the BBC. For the purpose of this analysis, the total charge 
in the BBC is divided into four centrality categories: 0\%--20\%, 
20\%--40\%, 40\%--60\%, and 60\%--86\%. The statistical significance of the data 
in the most peripheral bin (60\%--86\%) was too low to provide a useful 
measurement.  The MB trigger efficiency is 85.9 $\pm$ 2.0\% of the total 
\auau inelastic cross section at \sqsnsix. All MB data 
presented are calculated directly from the MB event sample.

The reaction-plane detector (RXNP) is a plastic scintillator paddle 
detector installed prior to the 2007 data-taking 
period~\cite{Richardson:2010hm}. It accurately measures the participant 
reaction-plane (RP) angle defined by the beam axis and the principal axis 
of the participant zone. The RXNP is located at $\pm$ 39 cm along the beam 
pipe from the center of PHENIX with a set of 24 scintillators in each arm.

In this paper, heavy-flavor hadrons are measured indirectly through 
electrons from the semi-leptonic decay channel. The two PHENIX central arm 
spectrometers (CA), which cover $|\eta|<0.35$ and $|\Delta\phi|=\pi/2$ each, 
provide track reconstruction, momentum and energy measurement, and 
electron identification ($e$ID) for this analysis. Based on the electron's 
bend in the magnetic field, the drift chambers and pad chambers 
reconstruct the track momentum with high resolution. The size 
and shape of the \v{C}erenkov ring detected in the ring imaging 
\v{C}erenkov detector (RICH) is used for electron identification over the 
full momentum range. Pions that fire the RICH above 5~GeV/$c$ are 
statistically insignificant. In addition, the electromagnetic calorimeter 
(EMCal)  measures the energy deposited by electrons and their shower 
shape. The energy-to-momentum ratio and the quality of matching of the 
shower shape to a particle template are used for $e$ID in a manner similar 
to the method used in~\cite{Adare:2010de}.
 
In the 2010 run the hadron-blind detector (HBD) was also installed in 
PHENIX~\cite{Anderson:2011jw}. The HBD is a windowless \v{C}erenkov 
detector that uses CF$_4$ gas as the radiator and amplification gas, in a 
container with a radius of $\approx$60~cm. The radiator is directly 
coupled in a windowless configuration to a readout element with a triple 
gas-electron-multiplier (GEM) stack.  The HBD is almost completely 
insensitive to hadrons up to around 4.5~GeV/$c$ when operated with a 
reverse-bias voltage, and therefore brings additional $e$ID capability. The 
HBD can also reduce background electrons from \pizero Dalitz decays and 
photon conversions in the detector material, especially conversions in the 
beampipe and entrance window into the HBD. A nearly field-free region in 
the HBD area (currents in the inner and outer coils of the central arm 
spectrometer magnets flow in opposite directions) preserves the opening 
angle of electron pairs and, given the large size of the readout pads, 
signals from a close pair will overlap on a cluster of neighboring pads. 
The $\pi^0$ Dalitz and conversion e$^+$e$^-$ pairs have small opening angles, 
and can therefore be rejected, while single electrons or electron pairs 
with large opening angles leave a signal of $\sim$ 20 photoelectrons 
(p.e.) in the HBD.

\section{Data Analysis}
\label{sec02DA}

\subsection{Candidate Electron Measurement \label{sec02_ince}}

To select data recorded with the optimum detector response, we 
use the average number of electrons and positrons per event in each run 
and reject those runs where the electron multiplicity deviates from the 
mean multiplicity by more than 3$\sigma$. To select good quality tracks, 
we follow the same method as described in~\cite{Adare:2010de}. The minimum 
transverse momentum for charged tracks in this analysis \pt is greater than 1.0 
GeV/$c$. For a track to be identified as an electron candidate, it is also 
required to fire the RICH and EMCal detectors, and to be associated with at 
least 4 fired phototubes in the RICH ring. In addition, the $E/p$ 
distribution, where $E$ is the energy deposited in the EMCal and $p$ is 
the momentum of the track reconstructed by the drift chambers, is used to select 
electron candidates. Electrons deposit most of their energy in the EMCal 
which makes $E/p$ close to 1, while hadrons deposit only part of the 
energy in EMCal which causes $E/p$ to be smaller. A cut of $dep > -2$ was 
used, where $dep=\frac{E/p-1}{\sigma_{E/p}}$.

In addition to the above electron cuts, the HBD provides electron 
identification and background rejection. We apply cuts on $hbdq$, where 
$hbdq$ is the number of p.e. recorded by the HBD in a 
cluster. Most of the hadrons and back plane conversion electrons are not 
associated with an HBD cluster; the rest can be divided into three 
categories:
  \begin{enumerate}
\item The track is a single electron (our signal) or an electron or 
positron from an electron pair with large opening angle; either case will 
produce an $hbdq$ distribution centered at 20 p.e.
\item The cluster comes from an electron pair with small opening angle 
which will produce an $hbdq$ distribution centered at 40 p.e.
\item The track does not fire the HBD itself, but is randomly associated 
with a fake cluster that is formed from the fluctuating HBD background. 
Charged particles traversing the CF$_4$ volume in the HBD produce 
scintillation light and creates hits with a small 
signal in random locations. In this case the $hbdq$ distribution has low 
values with an exponential shape. The minimum $hbdq$ cut removes 
most of these HBD background hits. A portion of these fluctuate to a 
larger $hbdq$ signal, but are statistically subtracted as described later 
in this section.
  \end{enumerate}

\begin{figure*}[htb]
 \includegraphics[width=0.998\linewidth]{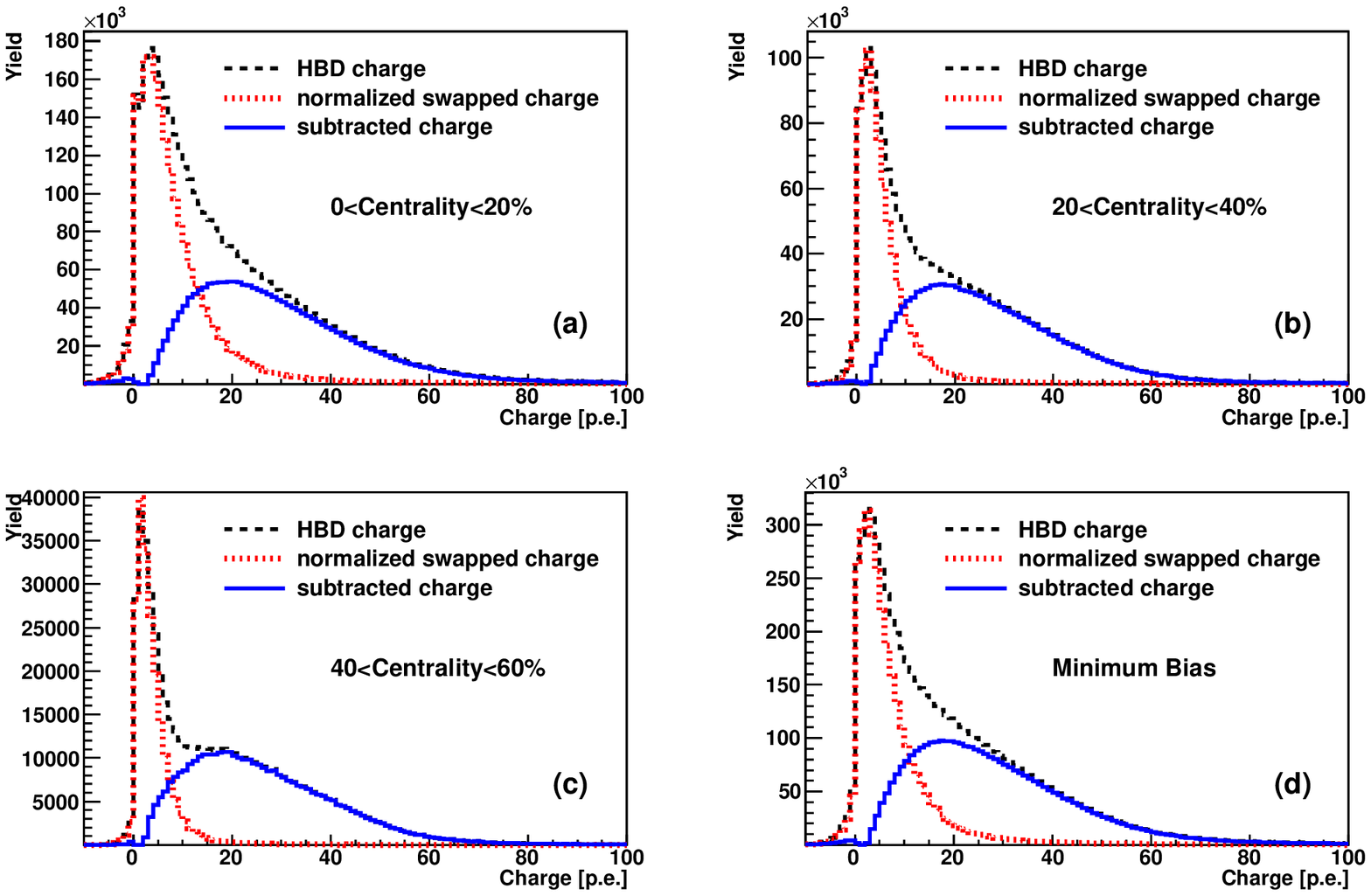}  
 \caption{\label{fig:hbdqcentrality}
(Color online) For centralities (a) 0\%--20\% (most central), 
(b) 20\%--40\%, 
(c) 40\%--60\%,
and (d) MB events, shown are the shapes of 
the HBD charge distribution (black dashed curves), the swapped HBD 
charge distribution (red dotted curves), and the subtracted HBD charge 
distribution (blue solid curves).  The swapped HBD distribution can 
statistically estimate the randomly matched HBD charges. 
}
 \includegraphics[width=0.998\linewidth]{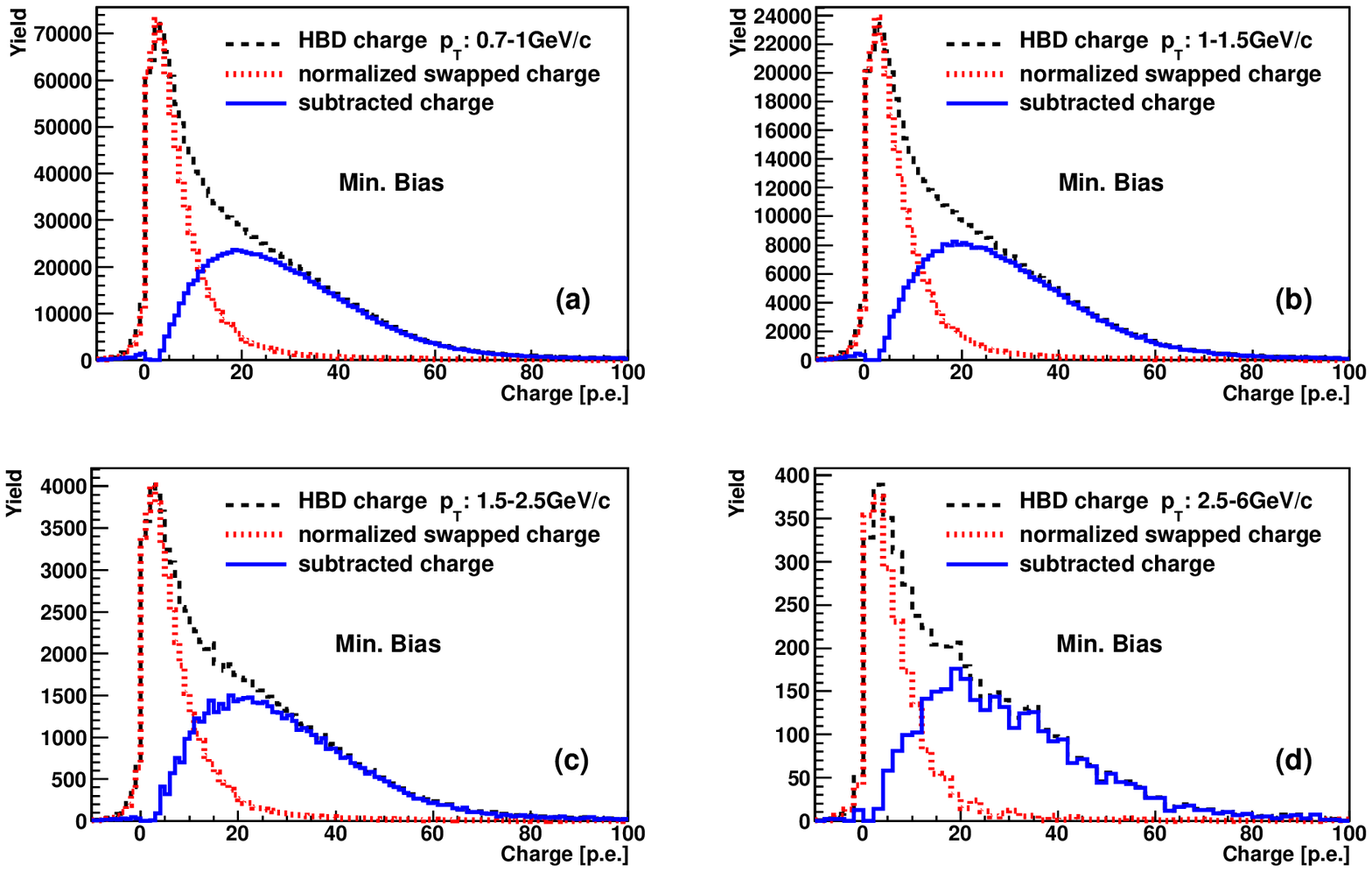}  
 \caption{\label{fig:hbdq_pt}
(Color online) The shape of the HBD charge distribution (black 
dashed curves), the swapped HBD charge distribution (red dotted curves) and 
the subtracted HBD charge distribution (blue solid curve) for the
indicated \pt ranges.  All plots are for MB events.
}
 \end{figure*}

\begin{figure*}[htb]
 \includegraphics[width=0.998\linewidth]{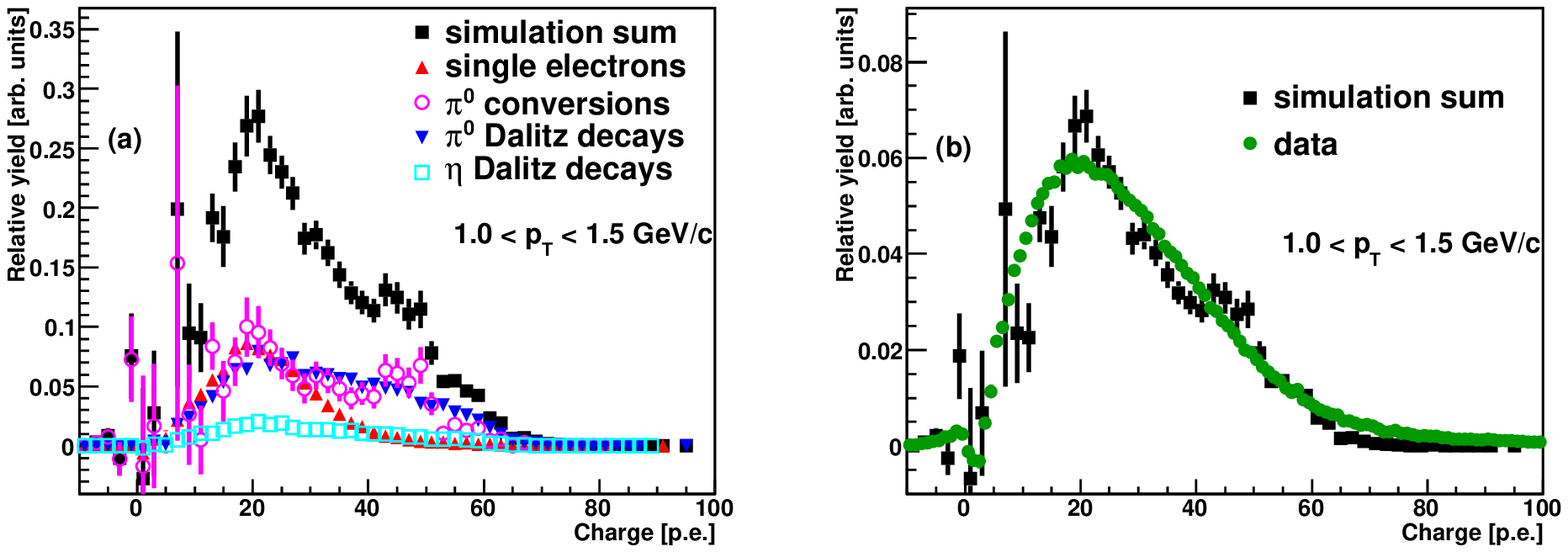}
 \includegraphics[width=0.998\linewidth]{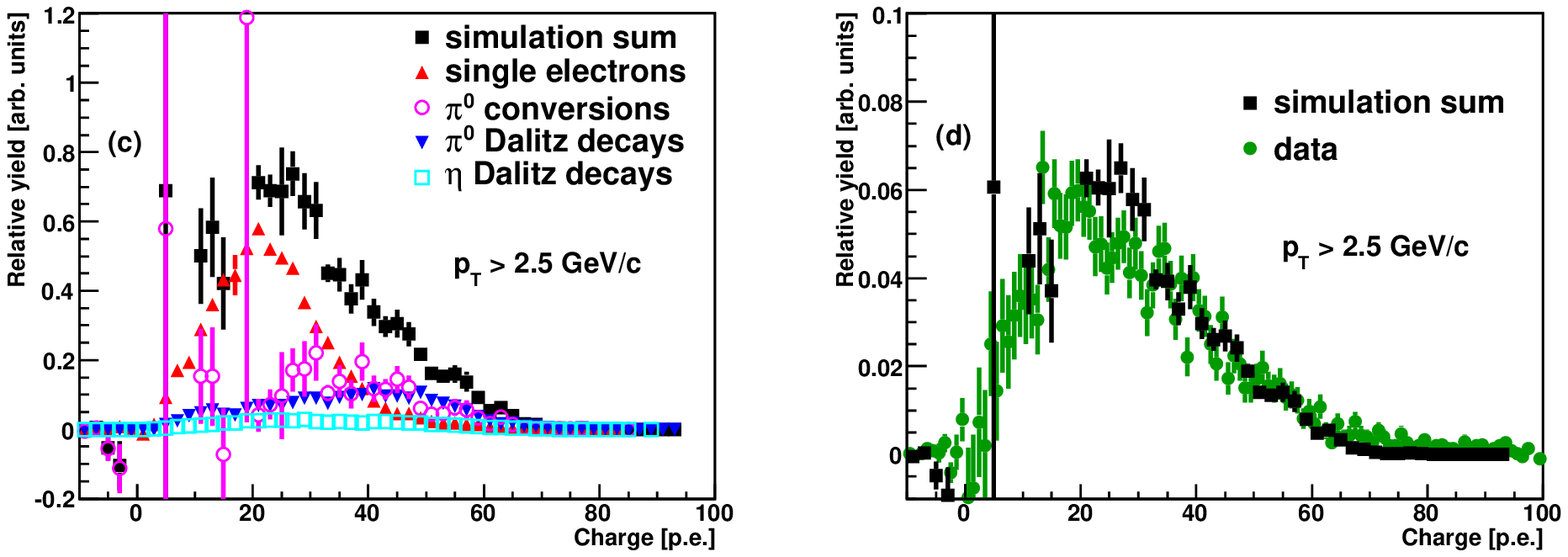}
 \caption{\label{fig:HBDPerformance}
(Color online) Simulated response of the HBD to different sources of electrons 
compared to the measured distribution for two different $p_T$ ranges, (a) 
$1.0<p_T<1.5$~GeV/$c$ and (c) $p_T>2.5$~GeV/$c$. (black squares) total 
simulation, (red triangles) single electrons, (blue inverted triangles) 
$\pi^0$ Dalitz decays, (open magenta circles) conversions, (open cyan 
squares) $\eta$ Dalitz decays, and (green circles) data.  For 
visual comparison, in (b) and (d) the distributions are normalized to 1
for the same \pt ranges as in (a) and (c).
} 
 \includegraphics[width=0.49\linewidth]{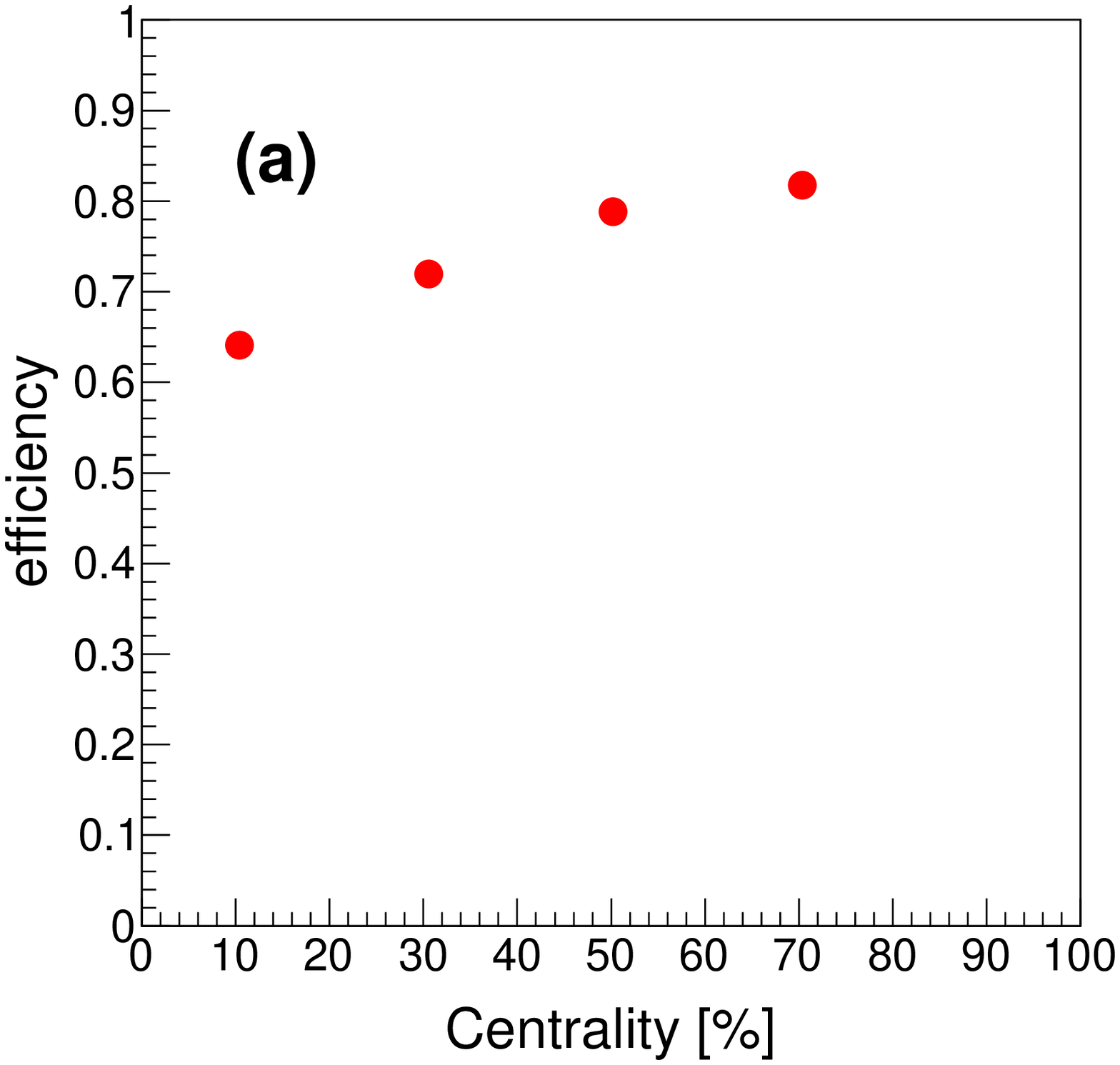}  
 \includegraphics[width=0.49\linewidth]{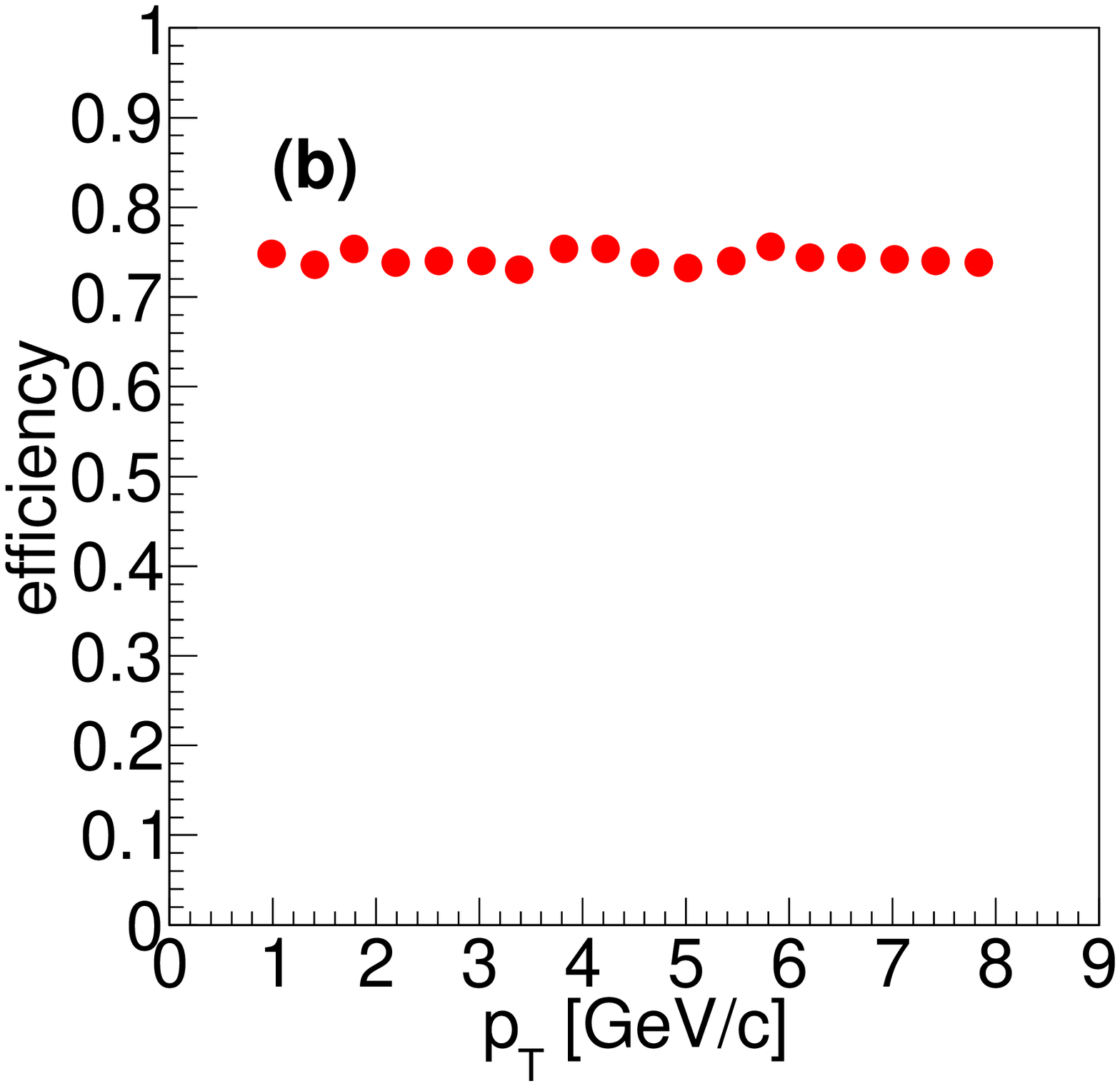}  
 \caption{\label{fig:embeddingEffHBD}
(Color online) The efficiency of the HBD cut, $10<hbdq<35$, for a single 
electron as a function of (a) centrality and (b) $p_T$. The efficiency was 
determined by embedding a simulated single-electron response into the  
real data.
}
 \end{figure*}

A cut of $10<hbdq<35$ reduces the backgrounds due to cases 2 and 3.  
A swapping method is used to statistically remove the background from case 
3, i.e. random track associations with HBD background, including 
conversions that are randomly associated with HBD clusters. The swapped 
HBD charge ($hbdq_s$) is obtained by matching in software a track found in 
the central arm to the HBD in the opposite arm, for example from HBD hits 
in the east arm to tracks in the west arm and vice versa. The swapped 
$hbdq$ distribution was normalized to the $hbdq$ in the bins near zero 
charge. Figure~\ref{fig:hbdqcentrality} shows the regular $hbdq$ 
distribution, the swapped $hbdq_s$ and the distribution after subtraction.

The swapped distribution, representing the $hbdq$ distribution for 
randomly associated tracks, falls rapidly. The swapped random coincidences 
produce signals at low $E/p$ as well as a peak centered at $E/p$ near 1. 
The low $E/p$ distribution is most probably random hadron coincidences and 
the peak is likely dominated by conversion electrons from the back plane of 
the HBD having a random coincidence with background clusters in the HBD. 
After subtracting the swapped distribution from the regular distribution, 
the $hbdq$ distribution has a peak around 20 p.e. and a long tail at high 
charge that is a superposition of the distribution of the single electron 
signal and the distribution of the close pair signal. To establish the 
extent of the remaining hadron contamination, the $dep$ distribution after 
subtraction is fit with a falling exponential (hadrons) and a Gaussian 
peaked close to 1 (electrons). The contamination changes with $p_T$ and 
centrality; it is largest at low $p_T$ and more central collisions. At 
$p_T$ near 1~GeV/$c$, the contamination is 2\% in the peripheral bin of 
40\%--60\%, 4\% in 20\%--40\% centrality collisions, and 8\% in 0\%--20\% central 
collisions. For $p_T>2$~GeV/$c$, the contamination is approximately 
independent of $p_T$ at 2\%, 2.5\%, and 3\% for the centrality bins 
40\%--60\%, 20\%--40\% and 0\%--20\% respectively. The yield within the range 
$10<hbdq<35$ of the swapped HBD charge distribution is subtracted at each 
\pt.
  
This swapping technique is repeated for each centrality and all 
distributions after subtraction are shown in Fig.~\ref{fig:hbdqcentrality}. In addition, the 
$hbdq$ distributions (subtracted or unsubtracted) broaden because of 
increasing fluctuations of the scintillation background in more central 
events. This will change the efficiency of the $hbdq$ cut as described in 
the subsection~\ref{sec02_simul}.
  
The distributions after subtraction are shown in Fig.~\ref{fig:hbdq_pt} for 
different $p_T$. The shape of the subtracted $hbdq$ distribution does not 
vary noticeably between 0.75 and 2.5~GeV/$c$. In central collisions, 
applying the $10<hbdq<35$ cut rejects 38\% of the tracks that satisfied 
the central arm's $e$ID selection. This fraction is 35\% for peripheral 
collisions. Some conversions still remain after the $hbdq$ cut and the 
swapping subtraction: these are subtracted using a simulated cocktail. The 
cocktail simulations are described in the subsection~\ref{sec02_bgsub}.

\subsection{Simulations}
\label{sec02_simul}

We use a {\sc geant}~\cite{Brun:1987ma} simulation to estimate the efficiency 
loss because of the inactive areas and the $e$ID cuts. This simulation has 
been demonstrated to match the central-arm PID and tracking-chamber 
performance as described in~\cite{Adare:2010de} and is used to 
determine the single-electron central-arm acceptance and efficiency. 
Because single electrons and close electron pairs have different $hbdq$ 
distributions, the efficiency of the HBD cut is different for electrons 
from different sources. Hence we use a cocktail of a variety of sources, 
the relative importance of which is constrained by available measured 
yields of different mesons.  Figure~\ref{fig:HBDPerformance} shows how 
well the HBD charge response is described by the HBD simulation.  The 
simulation has a bump at $hbdq \sim 45$, which 
is not observed in the data.


The HBD efficiency is 75\% for the single electrons in the simulation (and 
for electrons from pairs with very large opening angles). Within the 
simulation, we can examine which electron pairs are removed by the $hbdq$ 
cut.  This rejects 65\% of electrons that come from pairs that have a 
decay opening angle less than 0.05~radians while the rejection decreases 
until the opening angles reaches 0.1~radian. For each meson source in the 
cocktail, the efficiency is separately mapped as a function of $p_T$ and 
is used to correct the data.

We embed the simulated HBD single track response into real events to 
evaluate the centrality dependence of the HBD efficiency. For single 
electrons, the simulated $hbdq$ distribution is approximately Gaussian 
with a peak near 20. This broadens and shifts to a slightly higher average 
when embedded into a Au$+$Au event. The embedding efficiency for the fixed 
cut of $10<hbdq<35$ is calculated as a function of centrality and 
\pt. To understand the dependence of the efficiency on these two 
variables, we integrate over each in turn.  Figure~\ref{fig:embeddingEffHBD}(b) 
shows the \pt dependence integrated over 
centrality, which is approximately 75\% efficient independent of \pt. This 
lack of \pt dependence of the HBD cut efficiency is also observed in other 
centrality classes, but as seen in 
Fig.~\ref{fig:embeddingEffHBD}(a) the average efficiency does decrease for 
more central collisions; for central Au$+$Au events the efficiency has 
decreased to 65\%. As discussed earlier this is because of increased 
fluctuation of the underlying event background, mostly because of 
scintillation in the CF$_4$ gas.

\begin{figure}[htb]
\includegraphics[width=1.0\linewidth]{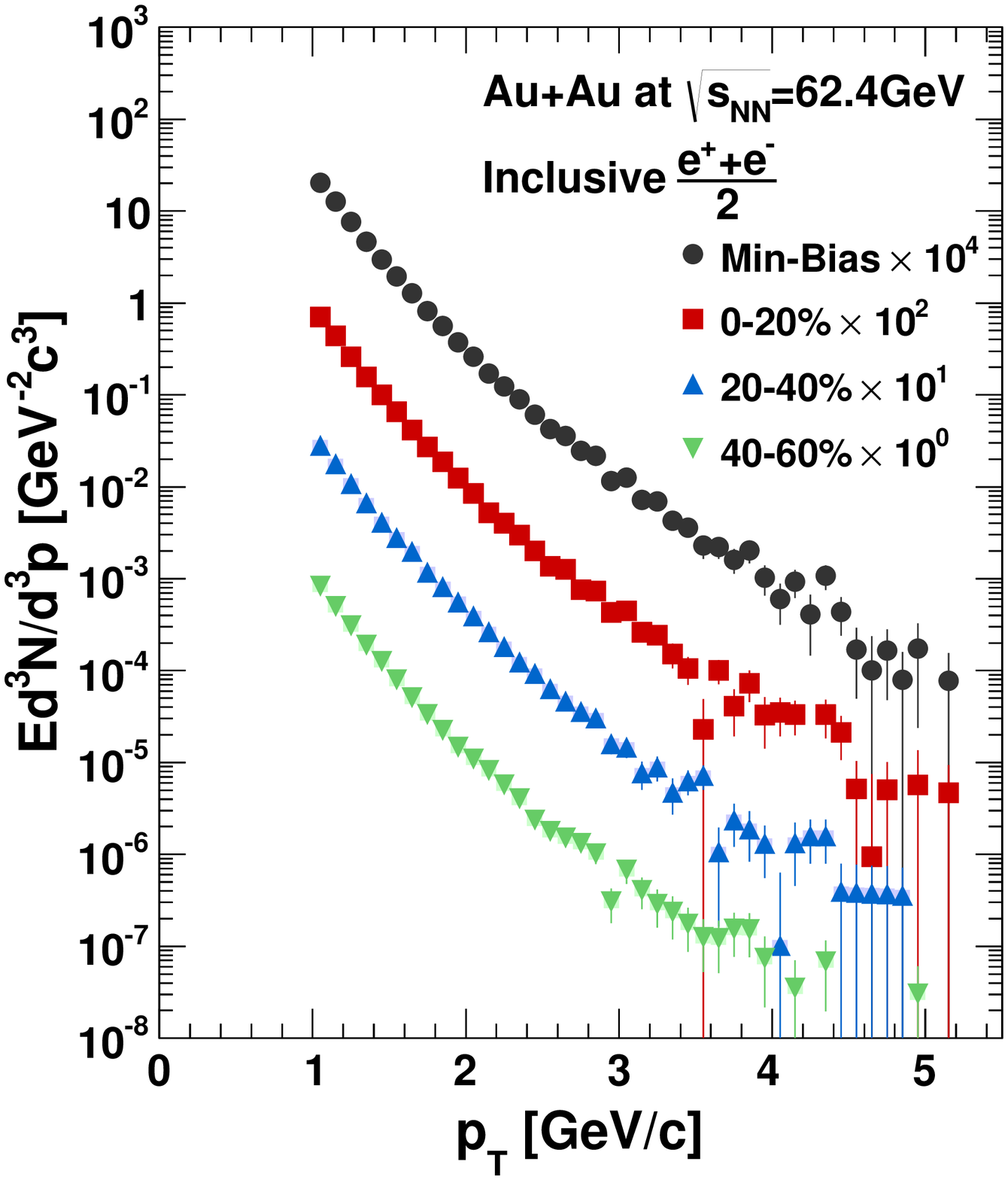}
 \caption{\label{fig:ince_yield}
(Color online) Invariant yield of candidate electrons measured in Au$+$Au 
collisions at \sqsn=62.4~GeV for different centrality bins. The yields 
are scaled by powers of 10 for clarity. The systematic uncertainty is 
shown as boxes and is, in many cases, comparable to the symbol size.
}
 \end{figure}

The acceptance and efficiency corrections are applied to the raw 
yields to produce the invariant yield of the electron candidates measured 
in Au$+$Au collisions at \sqsnsix for different centrality bins as shown 
in Fig.~\ref{fig:ince_yield}, where
\begin{equation}
E\frac{d^3N}{d^3p}=\frac{1}{2\pi p_T} 
\frac{1}{dy dp_T} \frac{1}{A \times \epsilon \times \epsilon_{\rm HBD}} 
\frac{N(e^+ + e^-)}{2} \frac{1}{N_{\rm events}},
\label{eq:invt}
\end{equation}
where $N(e^+ + e^-)$ is the number of electrons and positrons after HBD 
cuts, and after both swapped coincidences and hadron background 
contamination have been subtracted; $A{\times}\epsilon$ is the acceptance 
and efficiency of the central arm with $e$ID cuts, including embedding 
efficiency; and $\epsilon_{\rm HBD}$ is the efficiency of HBD cuts 
including embedding. In subsection~\ref{sec02_bgsub} a cocktail is used to 
subtract the remaining background statistically.


\subsection{Azimuthal anisotropy measurement of candidate electrons}
\label{sec02_azimuth}

For candidate electrons comprising photonic electrons and electrons from heavy 
flavor decay, we also measure the azimuthal anisotropy \vtwo, which is the 
second Fourier coefficient of the azimuthal distribution of the candidate 
electron yield with respect to the participant RP:
\begin{equation}
\frac{dN}{d\phi}=N_{0}(1+2v_{2}\cos 2(\phi- \Phi_{\rm RP})),
\label{eq:v2}
\end{equation}
where $\phi$ is the azimuthal angle of the electron track, $\Phi_{\rm RP}$ 
is the azimuthal angle of the participant RP, and $N_0$ is a 
normalization constant.

The participant RP is the plane formed by the transverse 
principal axis of the participants and the beam direction. The RXNP 
detector is used to measure the participant RP event by event. 
The event plane is constructed in two different windows:  
the South or North side of the RXNP. From these two planes we 
can calculate (Eq.~\ref{eq:RP_reso}) the RP resolution.
\begin{equation}
\langle\cos(2[{\Phi}_{\rm meas}-{\Phi}_{\rm real}])\rangle=\sqrt{2\langle\cos(2[{\Phi}^{S}_{m}-{\Phi}^{N}_{m}])\rangle},
\label{eq:RP_reso}
\end{equation}
where ${\Phi}^{S}_{m}$, ${\Phi}^{N}_{m}$ is the measured RP
using only South or North side of the detector. The RP
resolution is listed in Table~\ref{tab:ncoll} along with the number of 
binary collisions, \Ncoll, for each of the three centrality classes. 
\Ncoll was determined using a Glauber Monte Carlo calculation.

\begin{table}
\caption{\Ncoll values and RP resolution for each centrality class.}
\begin{ruledtabular}\begin{tabular}{ccc}
  centrality class & \Ncoll  & RP resolution \\ \hline
  0\%--20\% & 689.9$\pm$ 78.9  & 0.53 \\ 
  20\%--40\% & 270.5$\pm$ 27.5 & 0.62 \\ 
  40\%--60\% & 85.7$\pm$ 9.1  & 0.42 \\ 
\end{tabular}\end{ruledtabular}
\label{tab:ncoll}
\end{table}

\begin{figure*}[htb]
 \includegraphics[width=0.998\linewidth]{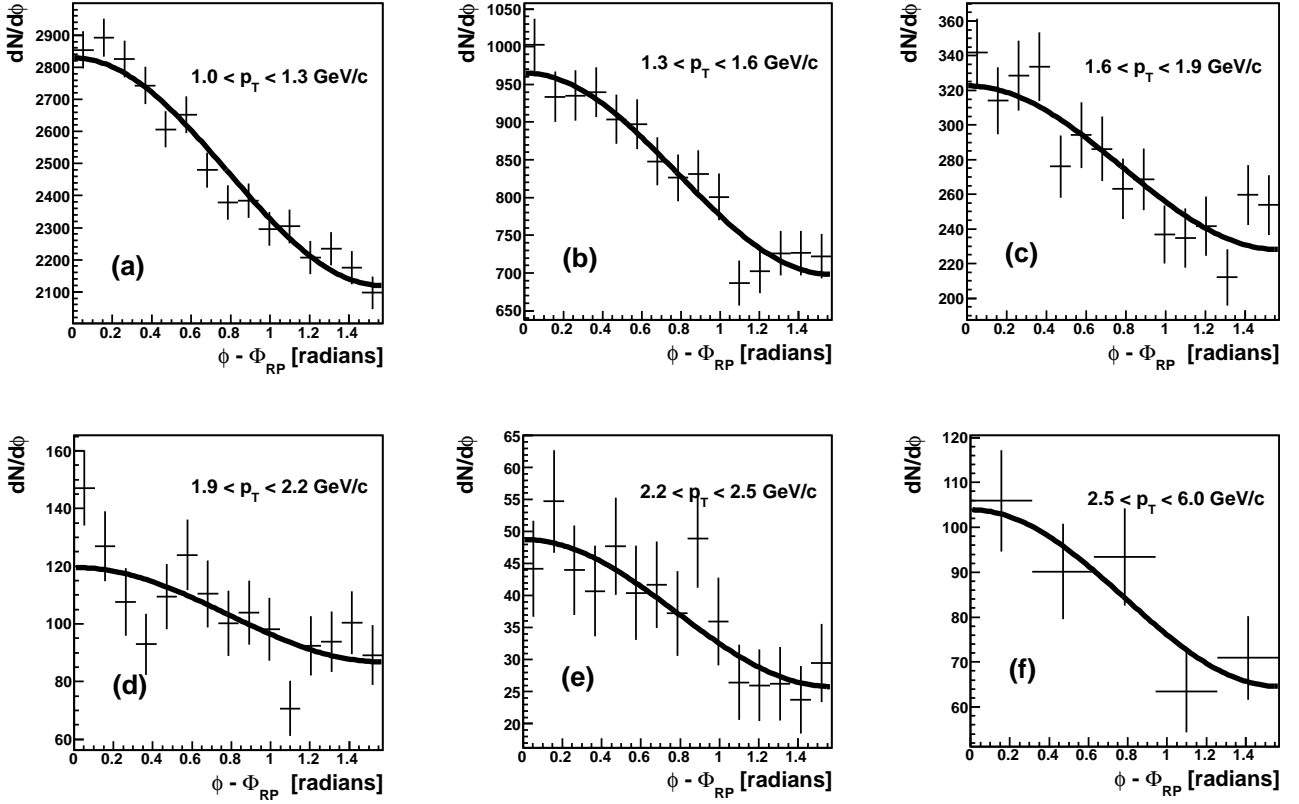}  
 \caption{\label{fig:v2_flow}
Candidate electron yield with respect to the RP for different 
\pt bins for events with centrality 20\%--40\% and fitted with the function 
$\frac{dN}{d\phi}=N_{0}(1+2v_{2}\cos 2(\phi- \Phi_{\rm RP}))$. The \pt 
bins are as indicated.
}
 \end{figure*}

\begin{figure}[htb]
\includegraphics[width=1.0\linewidth]{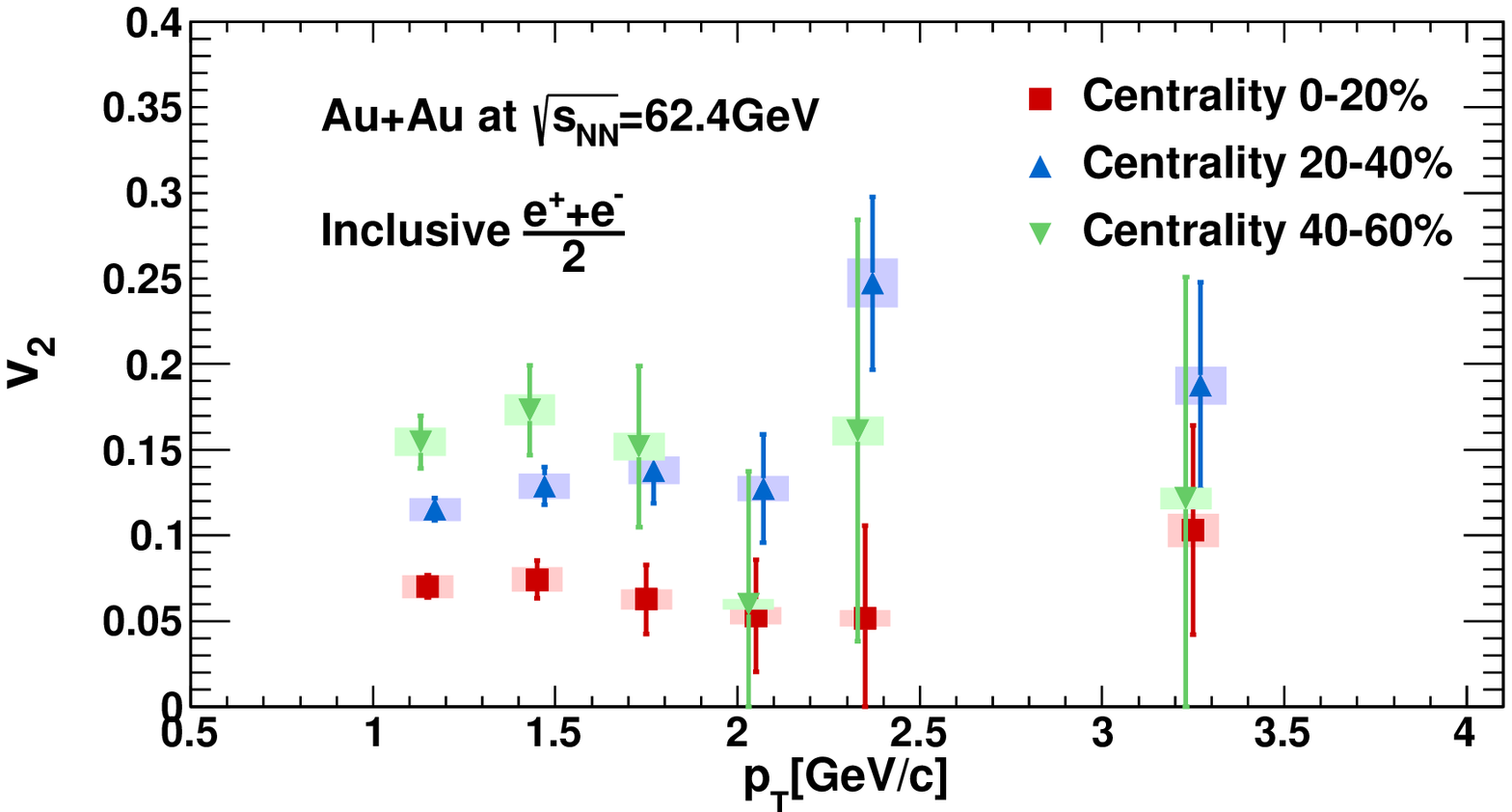}  
 \caption{\label{fig:ince_v2}
(Color online) Candidate electron $v_{2}$ as a function of \pt in Au$+$Au 
collisions at \sqsn=62.4~GeV for three different centrality bins. The 
systematic uncertainty is shown as boxes.
}
 \end{figure}


Figure~\ref{fig:v2_flow} shows the candidate electron yield with respect to 
the participant RP ($\phi- {\Phi}_{\rm RP}$) for selected \pt range for 
the 20\%--40\% centrality bin. The distribution is fitted with 
Eq.~\ref{eq:v2} to extract \vtworaw. By correcting the \vtworaw with the 
RP resolution (Eq.~\ref{eq:RP_reso_corr}), \vtwo of the 
particle distribution with respect to the real RP can be 
measured.
\begin{equation}
v_2=\frac{v_2^{\rm raw}}{\langle\cos(2[{\Phi}_{\rm meas}-{\Phi}_{\rm real}])\rangle},
\label{eq:RP_reso_corr}
\end{equation}
where $\Phi_{\rm meas}$ and $\Phi_{\rm real}$ are the measured and real 
RP angle. 
After correction by the RP resolution with 
Eq.~\ref{eq:RP_reso_corr}, the candidate electron \vtwo for different 
centrality bins is shown in Fig.~\ref{fig:ince_v2}.

\subsection{Cocktail Subtraction}
\label{sec02_bgsub}

As described above, the cut on $hbdq$ and the swapped subtraction removes 
most, but not all, of the background from photonic decays. In this section 
we describe the cocktail method of statistically subtracting the remaining 
electrons. A Monte Carlo event generator is used to produce electrons from 
hadron decays; the cocktail includes the photonic sources listed below:

\begin{itemize}

\item Dalitz decays of neutral mesons: $X \longrightarrow \gamma+e^-+e^+$, 
where $X=\pi^0,\eta,\eta',\rho,\omega,\phi$

\item Dilepton decays of neutral mesons: $X \longrightarrow e^-+e^+$, 
where $X=\rho,\omega,\phi$

\item Conversions of decay photons (including Dalitz) in detector material

\item $K_{e3}$ decays ($K \longrightarrow \pi^{\mp}+e^{\pm}+\nu_e^{(-)}$)

\item Conversion of direct photons

\end{itemize} 

\begin{figure}[htb]
\includegraphics[width=1.0\linewidth]{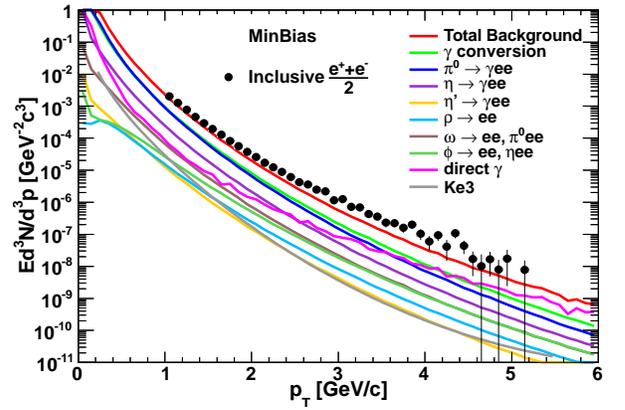}  
 \caption{\label{fig:cocktail}
(Color online) Invariant yield of (black dots) candidate electrons and 
(solid lines) electrons calculated from different photonic sources in 
Au$+$Au collisions at \sqsn$=$62.4~GeV for MB events.
}
 \end{figure}

The cocktail yield ($Y$) is calculated as
\begin{eqnarray}
Y&=&
\sum\epsilon_{\rm decay}({\rm hadron}, p_T)\times Y_{\rm decay}({\rm hadron}, p_T) \\  \nonumber
 &+&\sum \epsilon_{\rm conversion}(p_T){\times}R_{\rm CD}(p_T){\times}Y_{\rm Dalitz}({\rm hadron}, p_T) \\ \nonumber
 &+&\epsilon_{K_{e3}}(p_T)\times Y_{\rm decay}(K_{e3},p_T) \\ \nonumber
 &+&\epsilon_{\rm conversion}(p_T) \times Y_{\rm Conversion of direct photons}(p_T)
\label{eq:cocktail}
\end{eqnarray}
where $Y_{\rm decay}({\rm hadron}, p_T)$ is the yield of Dalitz and 
dilepton decays of neutral mesons. The efficiency and acceptance for each source are 
different as described in the subsection~\ref{sec02_simul}. For example, 
the efficiency for Dalitz decays of $\pi^0$ decreases from 0.5 at 
$p_T=1$~GeV/$c$ to 0.3 at $p_T=5$~GeV/$c$. 
Heavier mesons have larger opening angles and hence a higher probability 
for satisfying the HBD cuts.  For instance, $\eta$ decays have an 
efficiency of 0.6 at $p_T=1$~GeV/$c$ and 0.45 at $p_T=5$~GeV/$c$.
The conversion electrons are proportional to 
Dalitz decays with a proportionality factor $R_{\rm CD}$ based on simulation. 
$R_{\rm CD}$ is 0.9 at $p_T=1$~GeV/$c$ and increases linearly to to 1.4 at 
$p_T=5$~GeV/$c$.  This cocktail is constrained by the measured \pizero 
\pt spectra in \auau collisions at \sqsnsix~\cite{PhysRevLett.109.152301} which is 
fit to Eq.~\ref{eq:Hagedorn} for each centrality.

\begin{equation}
E\frac{d^{3}N}{d^{3}p_{T}}=\frac{c}{(e^{-ap_{T}-bp_{T}^2}+\frac{p_{T}}{p_{0}})^{n}},
\label{eq:Hagedorn}
\end{equation}
where $a$, $b$, $c$, $n$ and $p_0$ are fit parameters.  The relative 
normalization of other mesons to \pizero can be obtained from the meson to 
pion ratios at high 
\pt~\cite{Adler:2006hu,Riabov:2007sq,0954-3899-37-7A-075021}
\begin{itemize}
\item $\eta/\pi=0.48\pm0.03$
\item $\phi/\pi=1.00\pm0.30$
\item $\omega/\pi=0.90\pm0.06$
\item $\eta'/\pi=0.25\pm0.075$
\item $\eta/\pi=0.40\pm0.12$
\end{itemize}
and the shapes of the spectra assuming $m_T$ scaling, i.e. replace \pt 
with $m_T=\sqrt{\pt^{2}+{m_{\rm meson}}^{2}-{m_{\pizero}}^{2}}$ with the 
same parametrization of Eq.~\ref{eq:Hagedorn}. Figure~\ref{fig:cocktail} 
shows the cocktail of electrons from different photonic sources in Au$+$Au 
collisions at \sqsnsix for MB events. The electrons from photon 
conversions and from \pizero Dalitz decays are the largest contributions 
to the total cocktail background. The invariant yields of the candidate 
electrons are shown as black filled circles. There is more background from 
photon conversions in this measurement than in~\cite{Adare:2010de}.  This 
is the result of the removal of the helium bag and the installation of the 
HBD in the 2010 data taking, which increases the rate of photon 
conversions before the tracking detectors.  Most of the conversions that 
are removed using the cocktail are produced before the HBD, i.e. the 
beampipe, entrance window and gas. Only a very small portion (3\%) of the 
conversions subtracted using the cocktail come from the HBD itself.

The contribution from direct photons is significant for \pt~$>3$~GeV/$c$. 
For the contribution from direct photons, we use the measured 
\pt spectra from ISR R806, R807, R810 
experiments~\cite{0954-3899-23-7A-001} and \Ncoll scaling for each 
centrality bin. The electron spectra from $K_{e3}$ decays at \sqsnsix are 
obtained by a full {\sc geant} simulation of the PHENIX detector and the 
detector tracking algorithm. Because of a limited amount of experimental 
data on the $J/\psi$ \pt spectrum at midrapidity in Au$+$Au collisions at 
this energy, electrons from $J/\psi$ decays are not subtracted.  However, 
this background is small compared to the dominant backgrounds from pion 
decays and photon conversions.

The \vtwo of photonic electrons is calculated using a cocktail of sources. 
The PHENIX measurement of \vtwo of charged $\pi$ in Au$+$Au collisions 
at \sqsnsix~\cite{ppg124} 
is used to estimate 
the parent \pizero \vtwo distribution.  It is known that the measurements of 
the \vtwo of pions and kaons are the same as function of transverse 
kinetic energy~\cite{PhysRevC.85.064914},where transverse kinetic energy is 
$KE_T=\sqrt{p_T^2+m_o^2}-m_0$. Hence we assume that the \vtwo of other 
mesons in the cocktail have the same \vtwo values as a function of 
transverse kinetic energy as neutral pions. We assume the parent \vtwo is 
negligible for electrons from $K_{e3}$ decays and direct photons. The 
first background source is small for $p_T<3.5$~GeV/$c$ where we report \vtwo 
data. To account for possible flow of direct photons we increased the total 
systematic uncertainty of photonic \vtwo as described in the next 
subsection.

\begin{figure}[htb]
\includegraphics[width=1.0\linewidth]{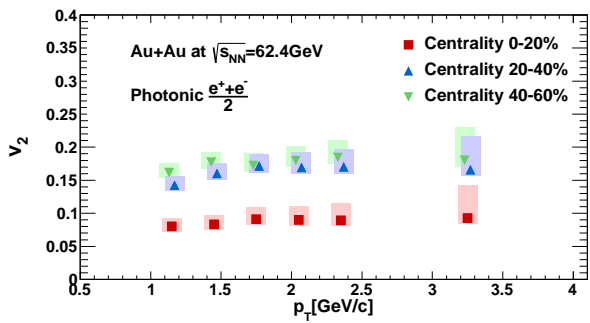}  
 \caption{\label{fig:phoe_v2}
(Color online) $v_{2}$ of photonic electrons calculated as the sum of 
different photonic sources in Au$+$Au collisions at \sqsn $=$ 62.4~GeV for 
three different centrality bins. Shaded boxes show the systematic uncertainties.
}
 \end{figure}

Figure~\ref{fig:phoe_v2} shows the estimated \vtwo of photonic electrons 
as a function of \pt for different centrality bins.

\subsection{Systematic Uncertainties\label{sec02_systematics}}

Systematic uncertainties in the candidate electron measurement include an 
overall 4\% contribution because of the acceptance. This was evaluated by 
calculating the difference in the geometrical matching between the 
simulation and the real data.  Other systematic uncertainties depend on 
\pt and are correlated.  For example, different choices in $e$ID cuts, loose 
and tight, were used to evaluate the systematic uncertainties due to $e$ID 
cuts. The variation between these sets is approximately independent of \pt 
at a level of 7\%. Alternative choices of HBD swapping normalization 
contribute 0.5\% to the systematic uncertainty, while different methods 
of selecting on HBD charge produced a \pt-dependent systematic uncertainty. 
The alternate cuts include changing the lower threshold of the $hbdq$ cut 
from 10 to 7 p.e., changing the upper cut from 35 p.e. to 30 or 40 p.e. 
These changes contribute a systematic uncertainty of 10\% for 
$p_T^e<1.5$~GeV/$c$ and a systematic uncertainty of 5\% for 
$1.5<p_T^e<6$~GeV/$c$.

Uncertainties in the cocktail method are mainly from the \pt-dependent 
uncertainties in the parent \pizero spectra which are taken from published 
data of \auau collisions at \sqsnsix~\cite{PhysRevLett.109.152301}. The 
uncertainties from the ratio of light mesons to pion yields are also 
extracted from published 
data~\cite{Adler:2006hu,Riabov:2007sq,0954-3899-37-7A-075021}.

We also assign a systematic uncertainty of 10\% for the amount of 
material in the {\sc geant} simulation used for the detector in the estimation 
of electrons from photon conversions, a systematic uncertainty of 20\% 
from the fits of direct photons and a conservative systematic uncertainty 
of 50\% for the electrons from $K_{e3}$ decays.

All these uncertainties are listed in Table~\ref{tab:systematicYield} and 
are propagated into the uncertainties of the heavy-flavor electron spectra 
by adding them in quadrature.

\begin{table}[htb]
 \caption{Systematic uncertainties in the yield of heavy-flavor electrons}
\begin{ruledtabular}\begin{tabular}{ll}
    Source & Description \\ \hline
    acceptance & 4\%  \\ 
    central arm $e$ID cuts & 7\%  \\ 
    HBD swapping & 0.5\%  \\ 
    HBD charge cut & $p_T$ dependent, 5\% to 10\%  \\ 
    cocktail & $p_T$ dependent, 10\% to 15\%   \\
    photon conversions & 10\% \\ 
    (in {\sc geant} material) & \\
    direct photon yield, & 20\% \\ 
    $K_{e3}$ & 0.25\% \\
\end{tabular}\end{ruledtabular}
 \label{tab:systematicYield}
\end{table}
\begin{figure}[htb]
\includegraphics[width=1.0\linewidth]{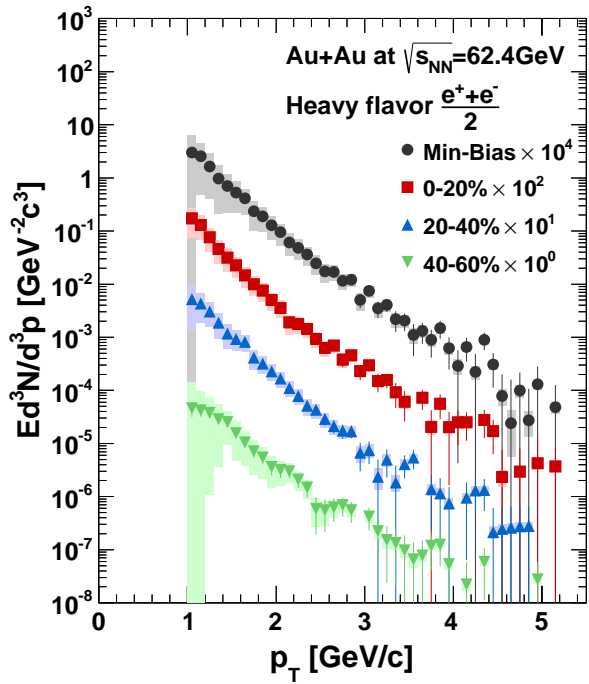}  
 \caption{\label{fig:hfe_yield}
(Color online) Invariant yield of heavy-flavor electrons measured in 
Au$+$Au collisions at \sqsnsix for different centrality bins. The yields are 
scaled by powers of 10 for clarity.   The uncertainty bars (boxes) show 
the statistical (systematic) uncertainties.
}
 \end{figure}
\begin{figure*}[hbt]
 \includegraphics[width=0.48\linewidth]{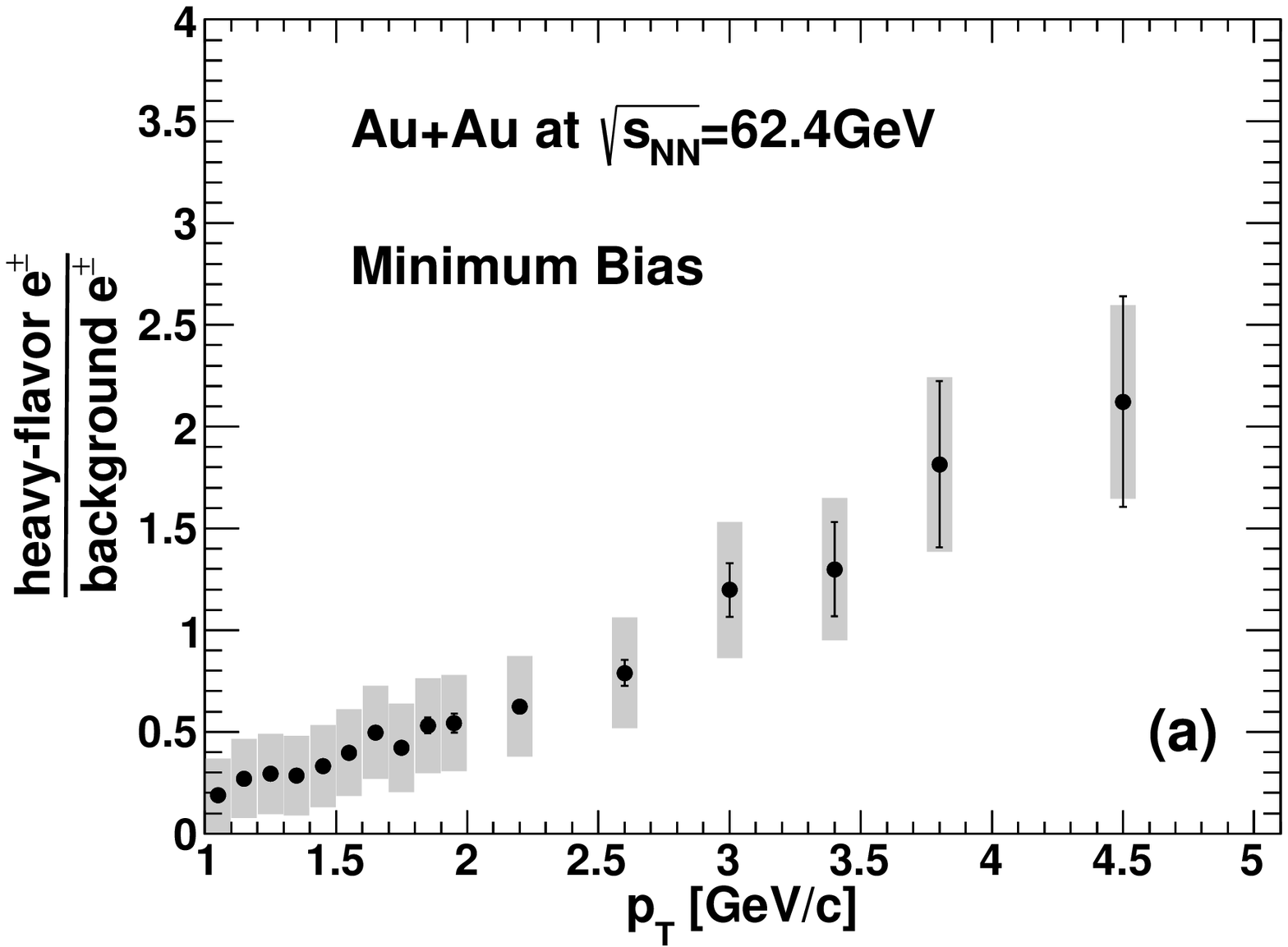}
 \includegraphics[width=0.48\linewidth]{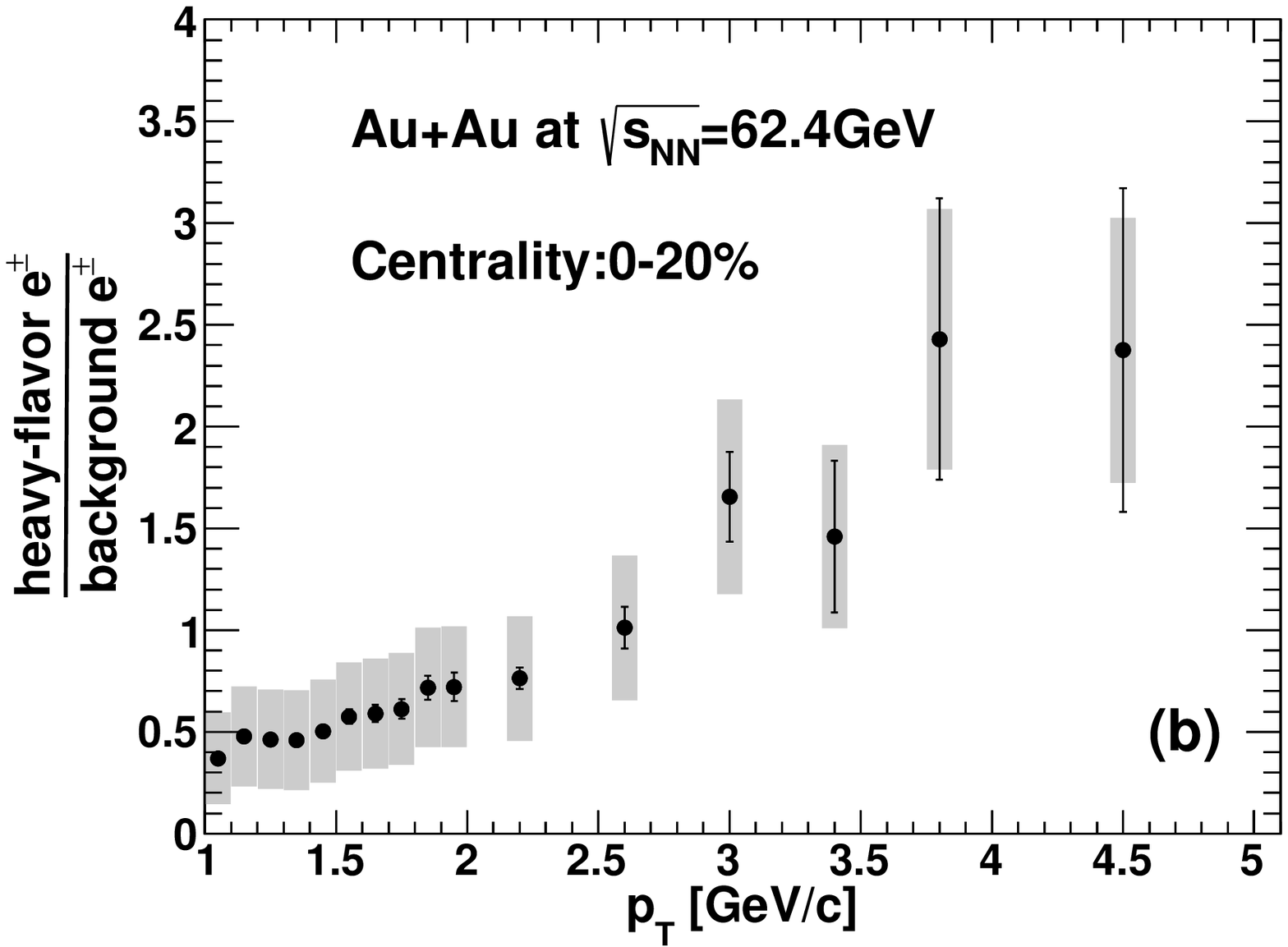}
 \includegraphics[width=0.48\linewidth]{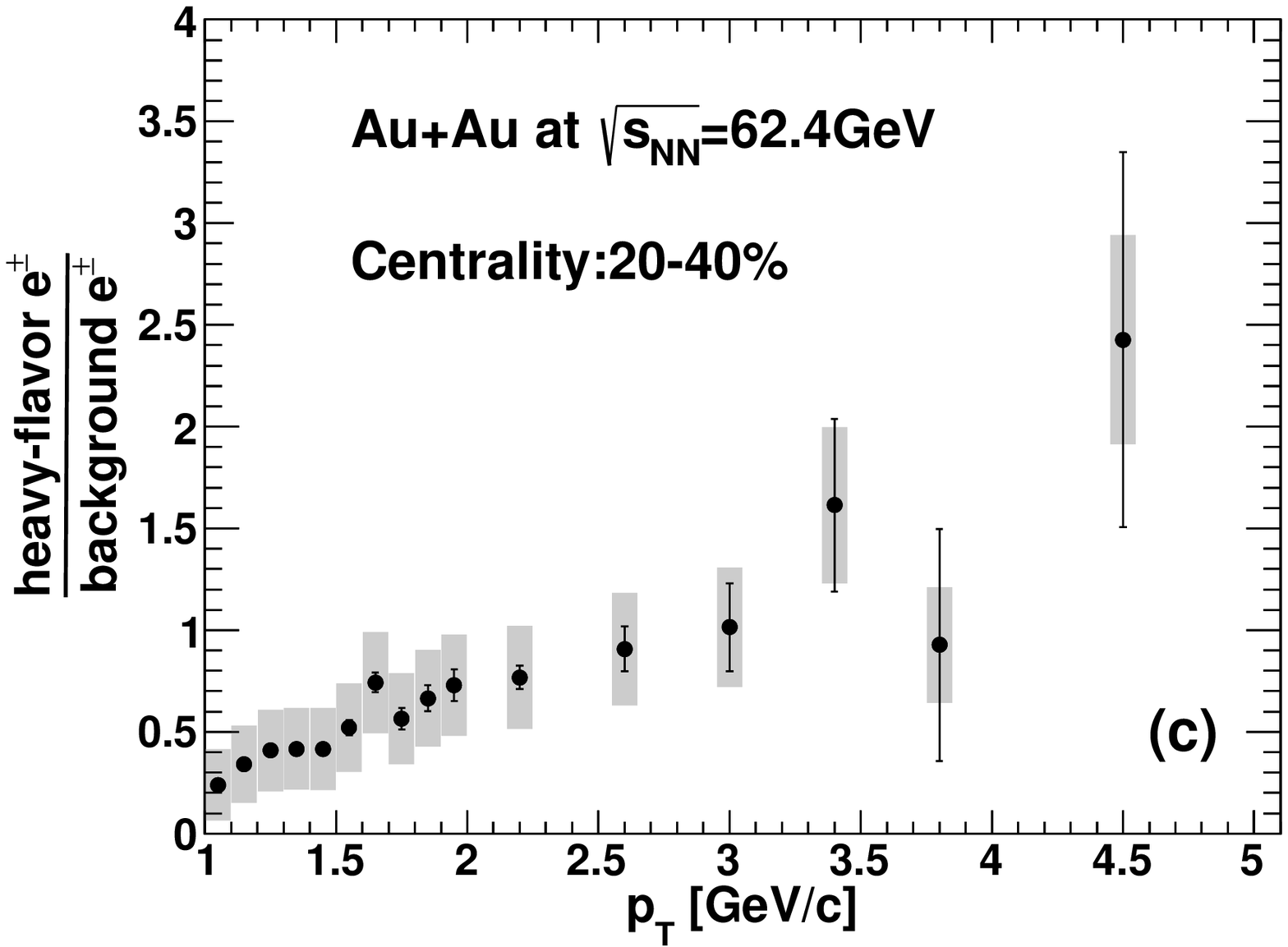}
 \includegraphics[width=0.48\linewidth]{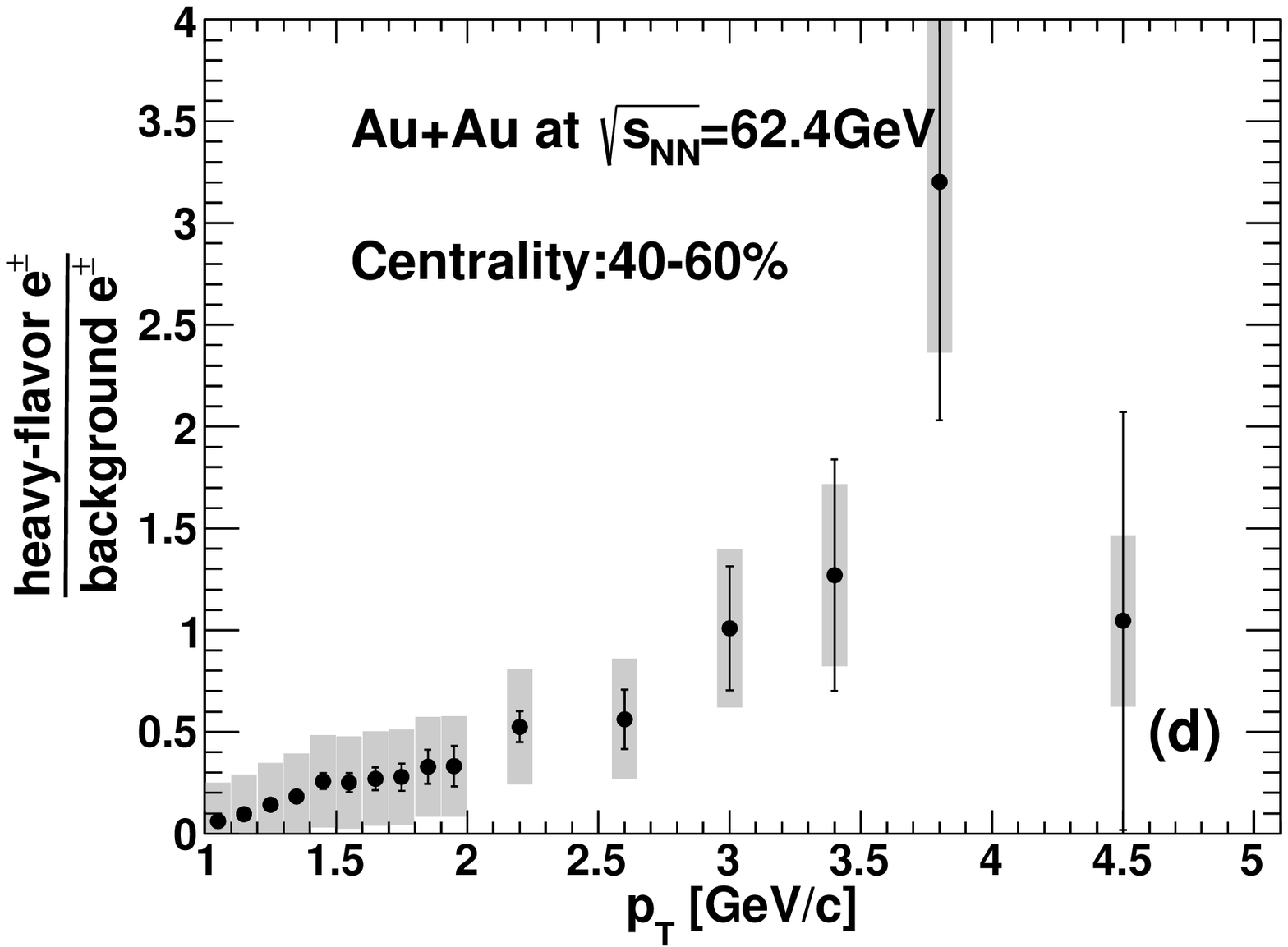}
 \caption{\label{fig:rnp_MB}
Ratio of the heavy-flavor electrons (signal) to photonic 
electrons (background) in Au$+$Au collisions at \sqsnsix for MB
events and the three indicated centrality classes that are used 
in this analysis.
}
 \end{figure*}

\begin{figure*}[htb]
   \begin{minipage}{0.6\linewidth}
\includegraphics[width=0.93\linewidth]{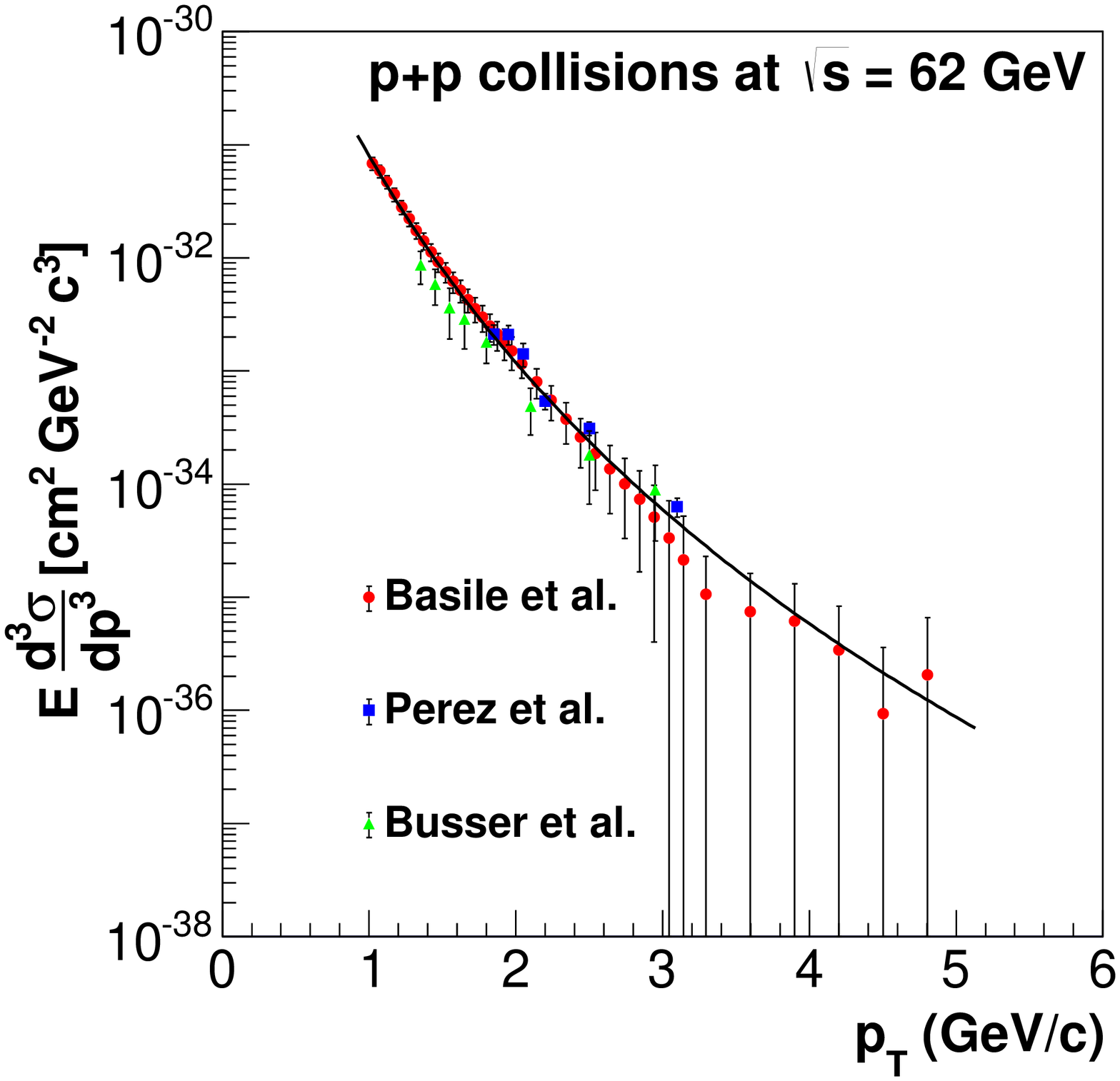}
   \end{minipage}
\hspace{0.5cm}
   \begin{minipage}{0.33\linewidth}
 \caption{\label{fig:pp}
(Color online) Invariant cross section of heavy-flavor electrons in $p$$+$$p$ 
collisions at \sqsnsix~\cite{Basile1981,Busser1976189,Perez1982260}. The 
curve is a combined power-law fit (see Table~\protect\ref{tab:pp}.
}
   \end{minipage}
\includegraphics[width=1.\linewidth]{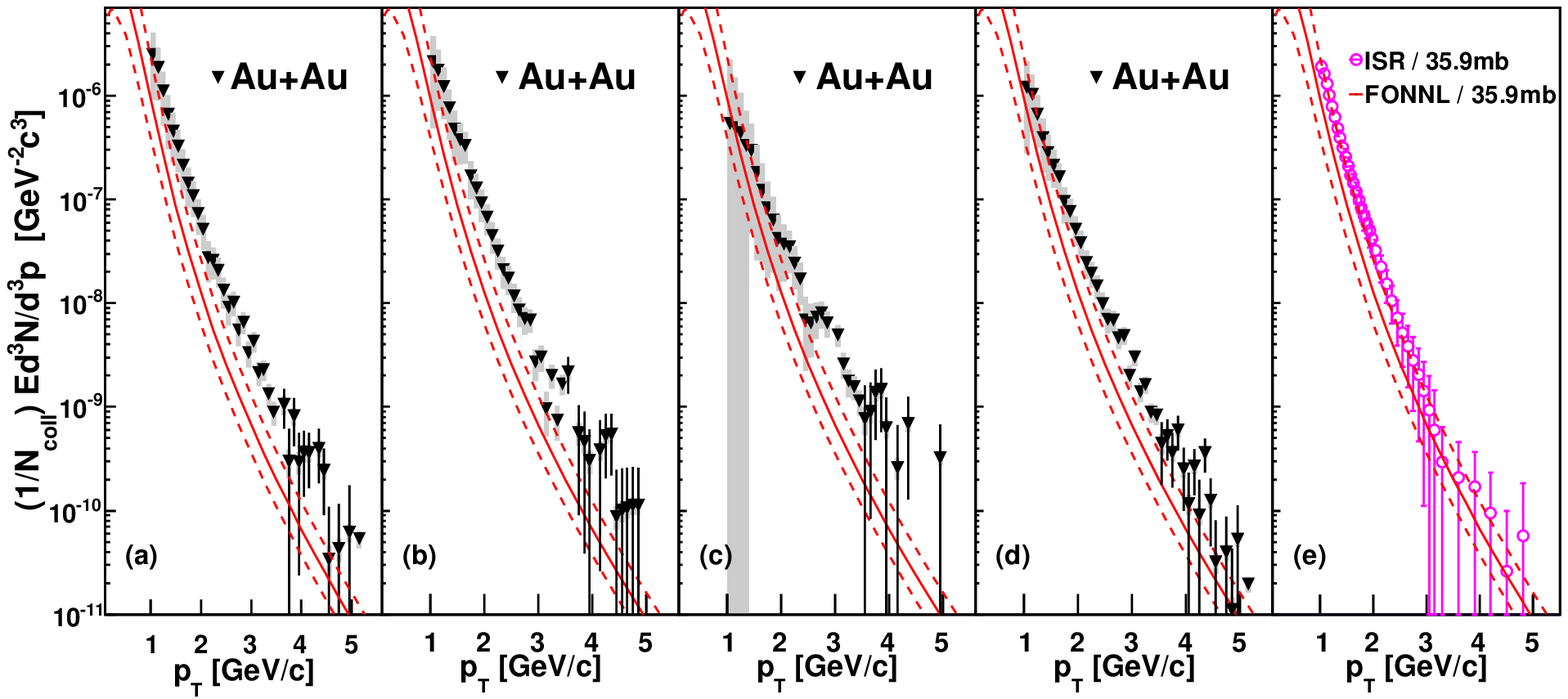}
 \caption{\label{fig:yield_ncoll}
(Color online) Invariant yield of heavy-flavor electrons per binary 
collisions in Au$+$Au collisions at \sqsnsix for (a) MB, 
(b) 0\%--20\%, (c) 20\%--40\%,  (d) 40\%--60\%, and 
(e) $p$$+$$p$ collisions at \sqsnsix measured in ISR experiments.
The curves are (red solid) FONLL calculations and (red dash) upper
and lower limits.  
}
 \end{figure*}

\begin{table}
\caption{Summary of fit characteristics for the 
invariant cross section of heavy-flavor electrons in $p$$+$$p$
collisions at \sqsnsix~\cite{Basile1981,Busser1976189,Perez1982260},
as shown in Fig.~\protect\ref{fig:pp}}.
\begin{ruledtabular}\begin{tabular}{ccc}
 $p_{T}$ (GeV/$c$) & Fit Value (cm$^{2}$GeV$^{-2}c^{3}$) 
 & Relative Uncertainty\\
 \hline
1.1 & 4.82$\times10^{-32}$ & 0.061\\
1.3 & 1.88$\times10^{-32}$ & 0.038\\
1.5 & 7.89$\times10^{-33}$ & 0.038\\
1.7 & 3.55$\times10^{-33}$ & 0.045\\
1.9 & 1.69$\times10^{-33}$ & 0.049\\
2.1 & 8.44$\times10^{-34}$ & 0.052\\
2.3 & 4.40$\times10^{-34}$ & 0.055\\
2.5 & 2.39$\times10^{-34}$ & 0.061\\
2.7 & 1.34$\times10^{-34}$ & 0.070\\
2.9 & 7.74$\times10^{-35}$ & 0.083\\
3.1 & 4.60$\times10^{-35}$ & 0.10\\
3.4 & 2.21$\times10^{-35}$ & 0.13\\
3.8 & 8.92$\times10^{-36}$ & 0.17\\
4.2 & 3.87$\times10^{-36}$ & 0.219\\
4.6 & 1.79$\times10^{-36}$ & 0.27\\
5.0 & 8.72$\times10^{-37}$ & 0.32\\
\end{tabular}\end{ruledtabular}
\label{tab:pp}
\end{table}

The systematic uncertainties on the \vtwo measurement include the 
uncertainty in electron candidate \vtwo and the uncertainty in the 
photonic electron \vtwo. The uncertainty in electron candidate \vtwo is 
because of the RP resolution (5\%). The systematic uncertainty 
is 8\% for central \vtwo and 5\% for midcentral photonic electrons.  We 
find a systematic uncertainty of 4\% due to the uncertainties of the 
relative ratio of different photonic-electron sources to the photonic 
electron \vtwo.

We also assign an additional systematic uncertainty because of possible 
flow of direct photons as observed in Au$+$Au collisions at 
\sqsn=200~GeV~\cite{PhysRevLett.109.122302}, 
which was assumed to 
be zero in our calculation of photonic flow. This additional systematic 
uncertainty was calculated assuming that direct photon flow is the same as 
that of \pizero.

\section{Results and Discussion}
\label{sec03}

\subsection{Heavy Flavor Electron Yield}

To extract the invariant yield of heavy-flavor electrons, the photonic 
electron background is subtracted from the invariant yield of candidate 
electrons for each centrality bin.
\begin{equation}
E\frac{d^3N_{\rm heavy flavor}}{d^3p}=E\frac{d^3N_{\rm inclusive}}{d^3p}-E\frac{d^3N_{\rm cocktail}}{d^3p}, 
\end{equation}
i.e. the data shown in Fig~\ref{fig:ince_yield} minus the 
centrality-dependent cocktail comparable to Fig~\ref{fig:cocktail}.  
Figure~\ref{fig:hfe_yield} shows the invariant yield of heavy flavor 
electrons as a function of \pt in four different centrality ranges, 
MB, 0 to 20\%, 20\%--40\%, and 40\%--60\%. The error bars and error 
boxes represent respectively the statistical and systematic uncertainties 
in the heavy-flavor electron measurement.


Figure~\ref{fig:rnp_MB} shows the signal to background ratio \rsb 
(Eq.~\ref{eq:rnp_eq0}), in MB events and for the three 
centrality classes used in this analysis.
\begin{equation}
\rsb=\frac{N_{\rm hf}}{N_{\rm photonic}},
\label{eq:rnp_eq0}
\end{equation}
where $N_{\rm hf}$ is the yield of heavy-flavor electrons, $N_{photonic}$ is 
the yield of photonic electrons, i.e. the data shown in 
Fig~\ref{fig:ince_yield} divided by the centrality-dependent cocktail 
comparable to Fig~\ref{fig:cocktail}. \rsb increases with \pt. At low \pt 
the candidate electrons are primarily from the photonic sources. At high 
\pt, electrons from heavy flavor meson decays start to dominate the 
candidate electron yield.


As a baseline, there are three available $p$$+$$p$ results from the 
ISR~\cite{Basile1981,Busser1976189,Perez1982260} that are shown in 
Fig.~\ref{fig:pp}. 
Table~\ref{tab:pp} shows the value of the fit and its relative 
uncertainty for each $p_{T}$ point used to calculate $R_{AA}$.
These data sets are simultaneously fit to a power-law function:
\begin{equation}
yield= \frac{a}{(p_T+b)^n},
\end{equation}
where the parameters are determined to be $a=1.21 \pm 3.55 \times
10^{-28}$, $b=1.015\pm0.39$~GeV/$c$, and $n=10.45\pm1.43$, as shown in
Fig.~\ref{fig:pp}.

\begin{figure*}[htb]
 \includegraphics[width=0.325\linewidth]{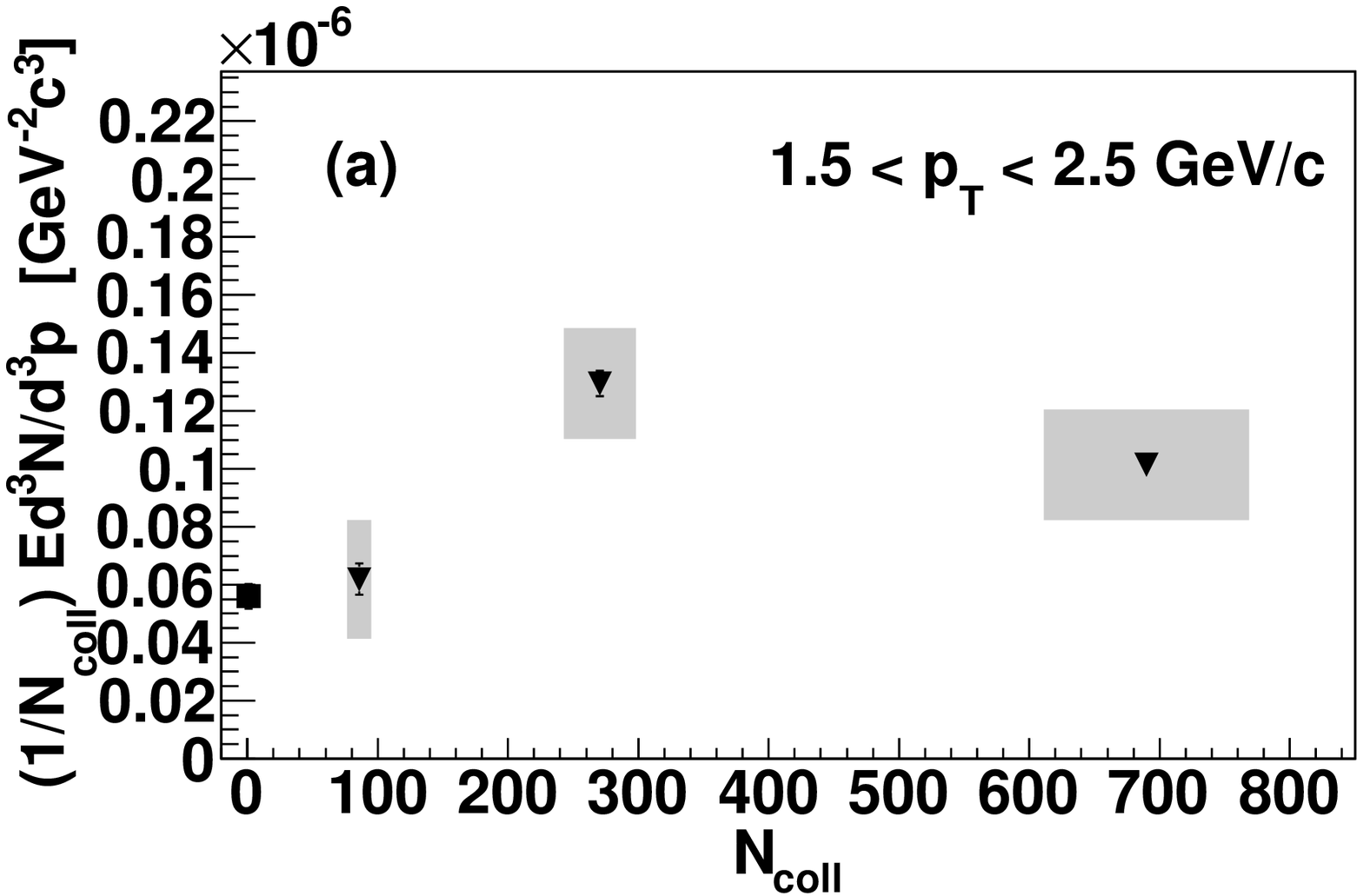}
 \includegraphics[width=0.325\linewidth]{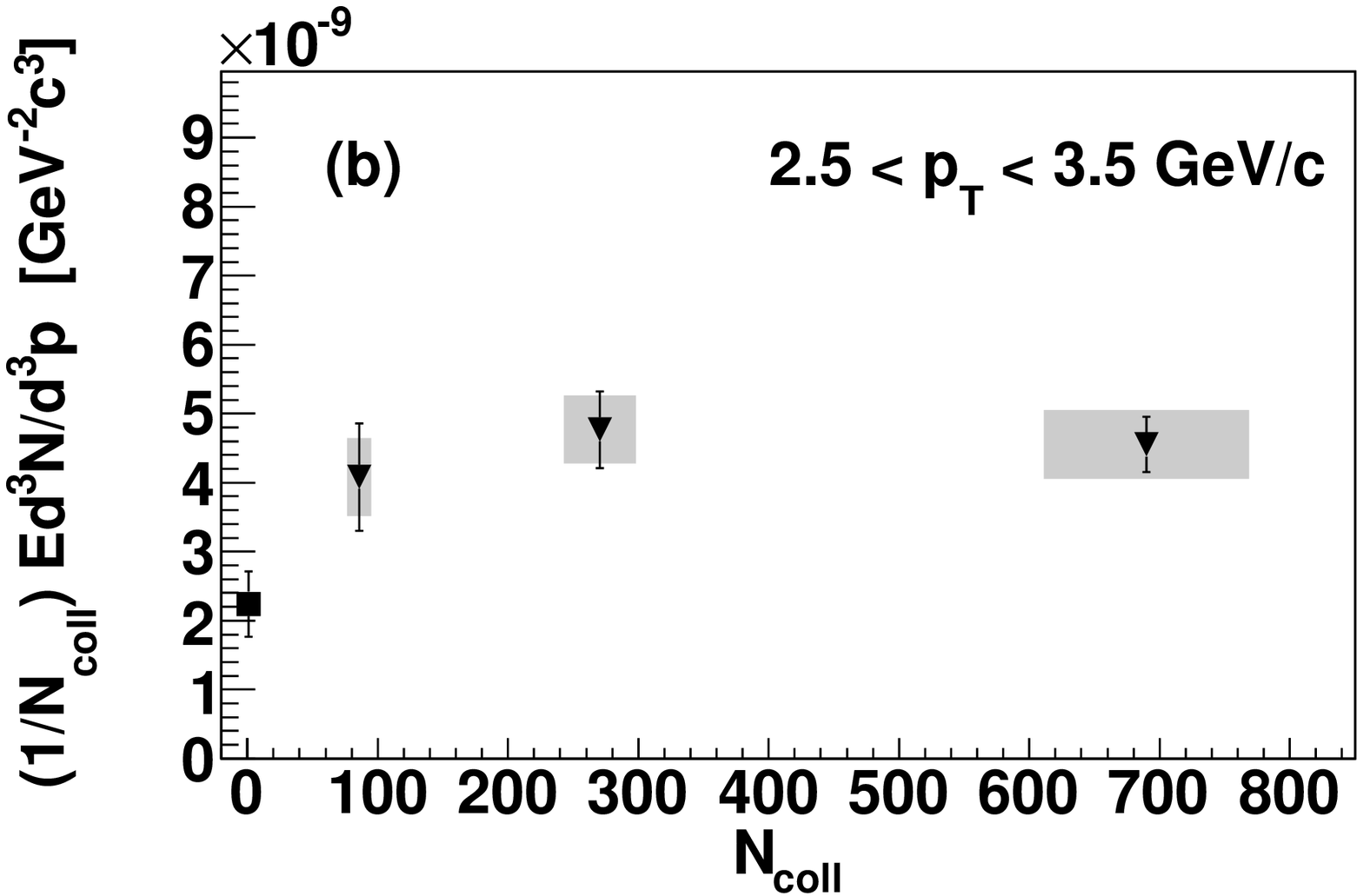}
 \includegraphics[width=0.325\linewidth]{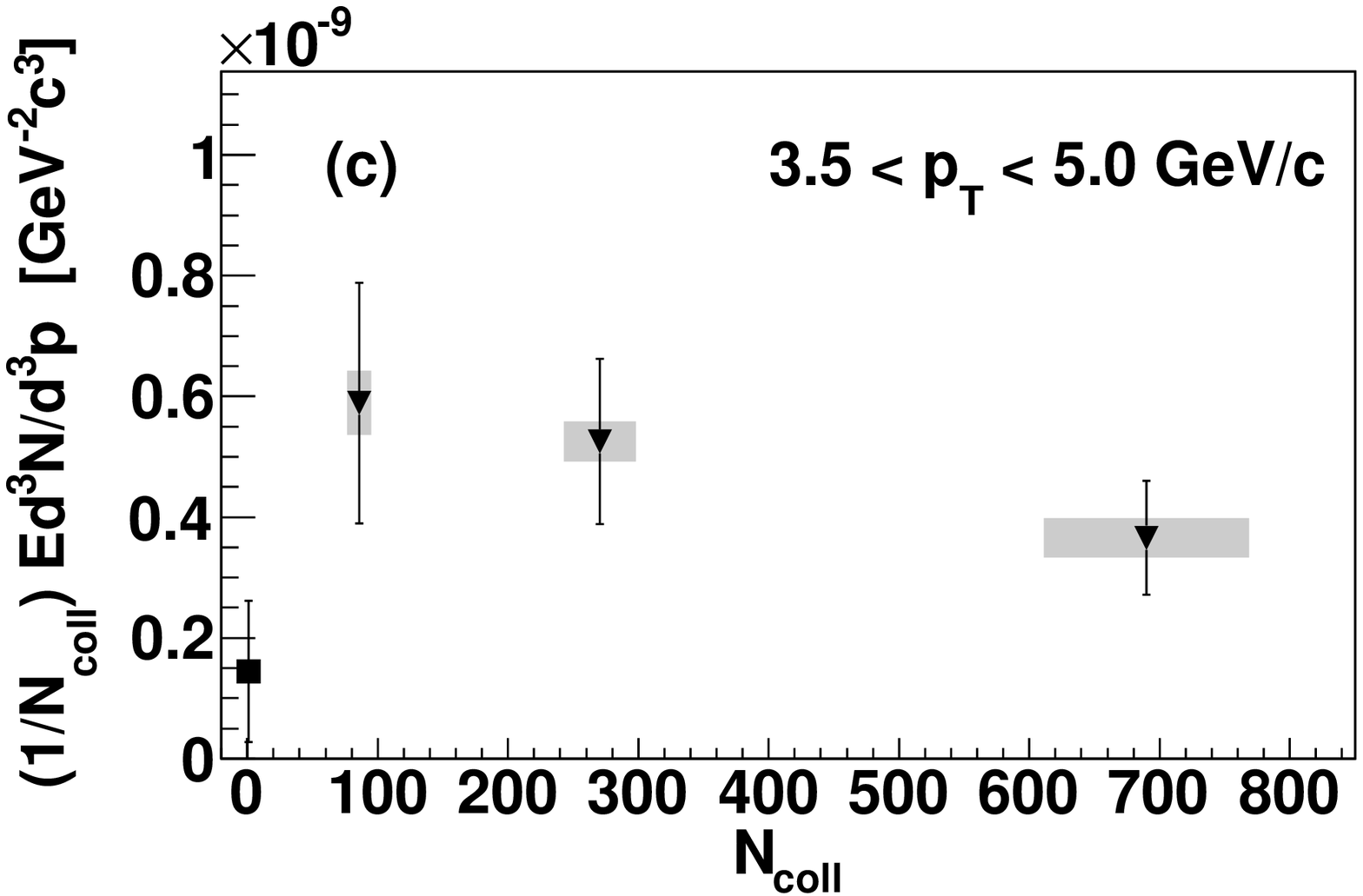}
\caption{\label{fig:integrat_yield_only}
Integrated invariant yield per binary collision vs \Ncoll 
for heavy-flavor electrons in Au$+$Au collisions at \sqsnsix
for the indicated \pte ranges.  The data point at \Ncoll$=$1
is for \pp collisions.
}
  \includegraphics[width=0.45\linewidth]{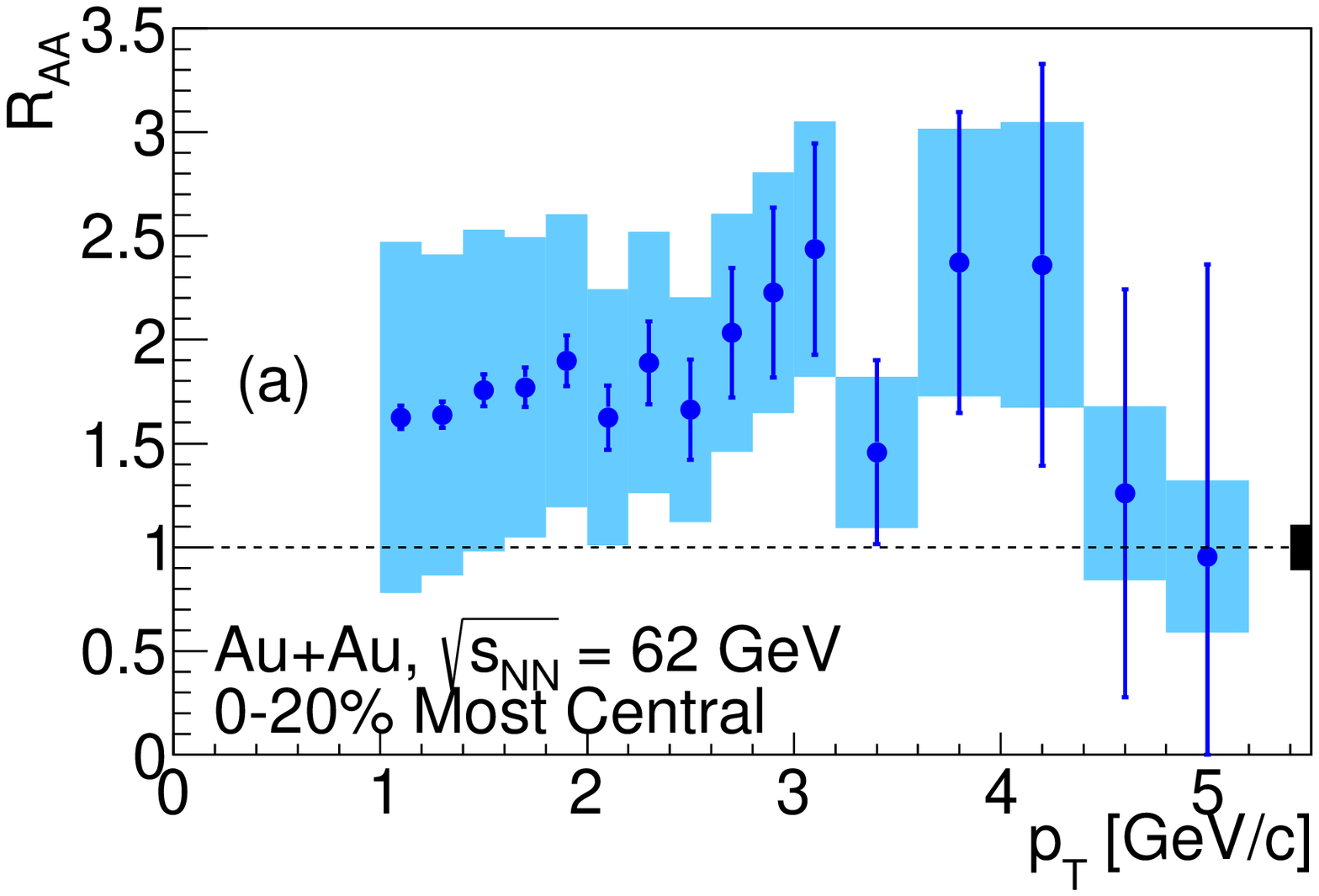}
  \includegraphics[width=0.45\linewidth]{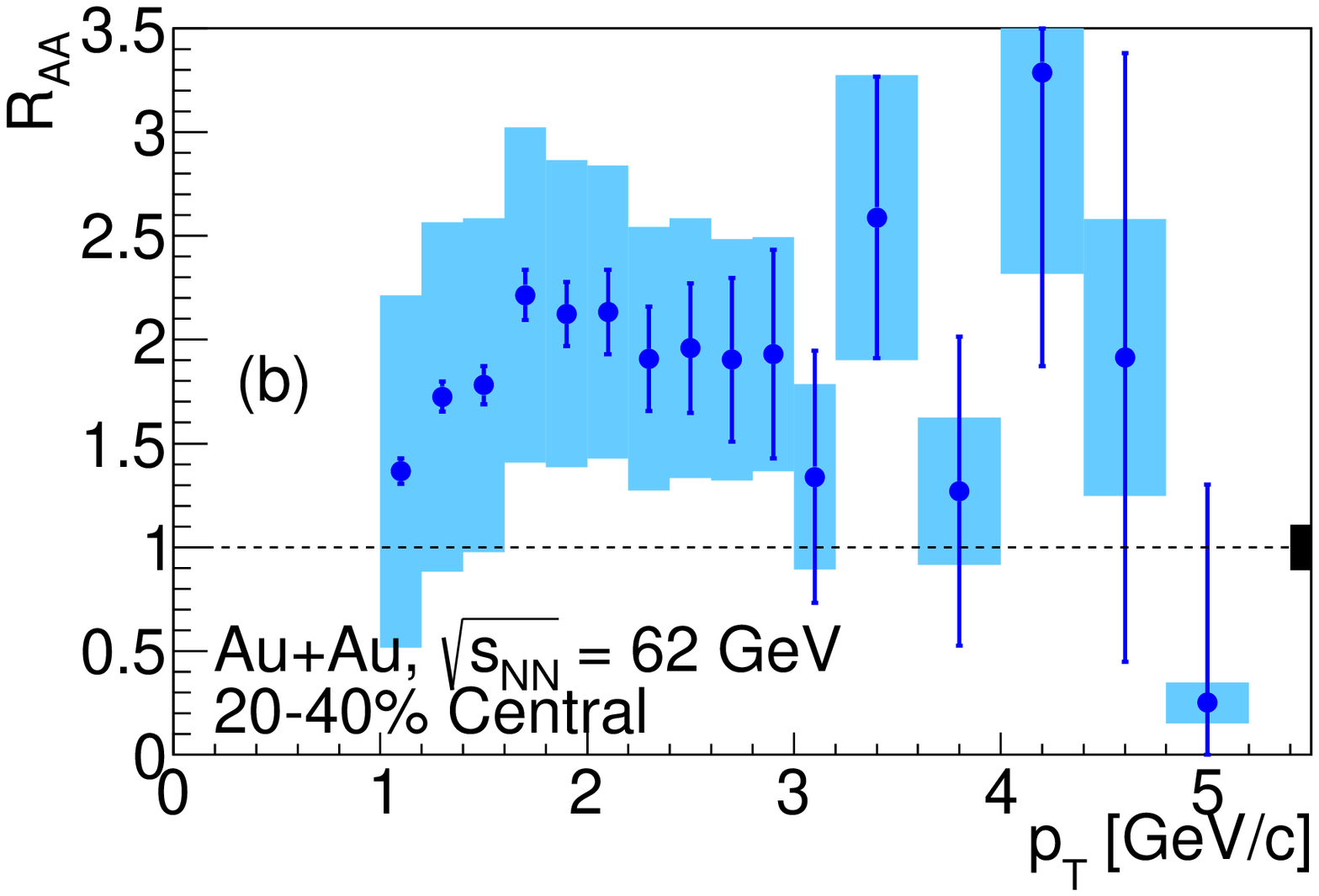}
  \includegraphics[width=0.45\linewidth]{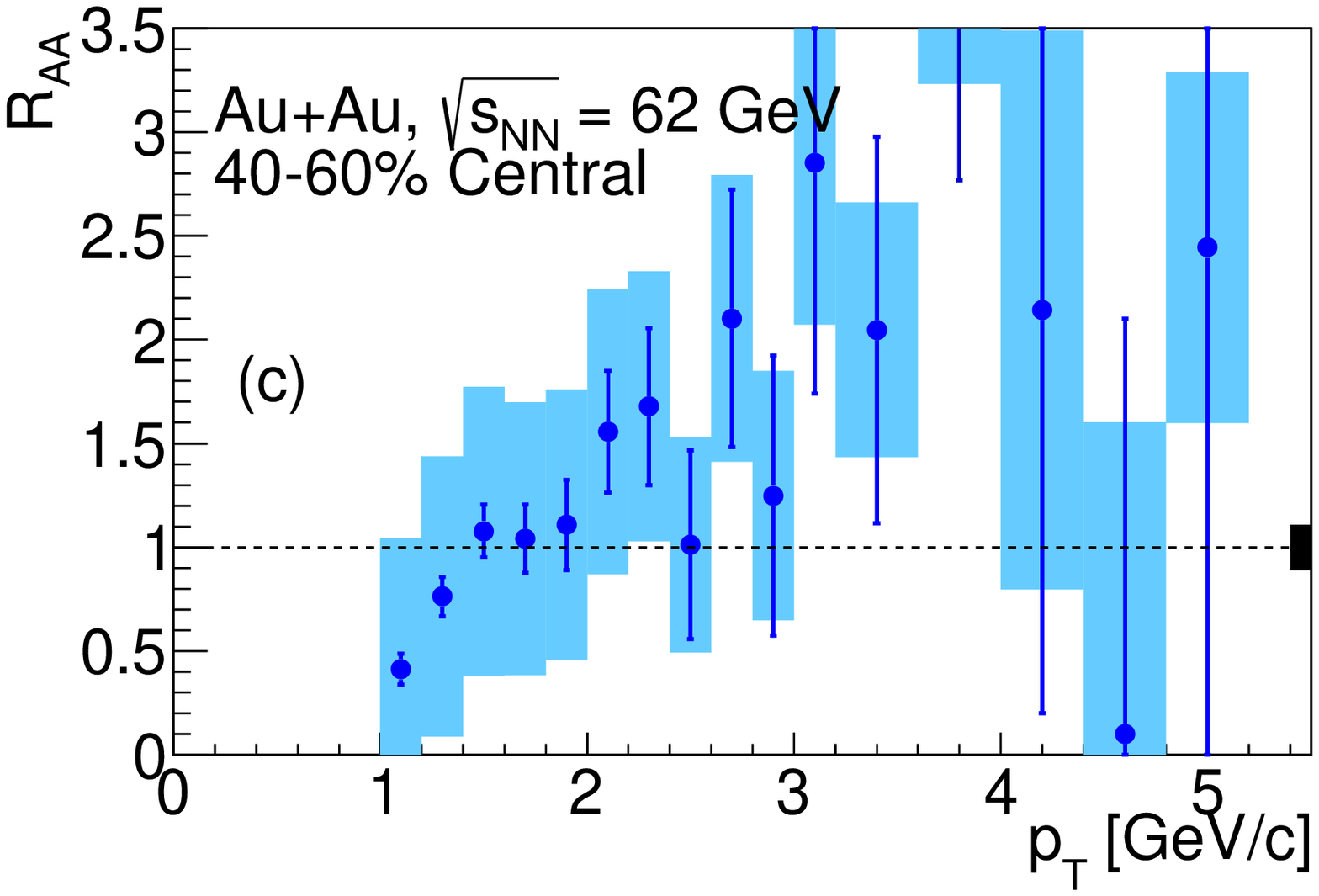}
  \includegraphics[width=0.45\linewidth]{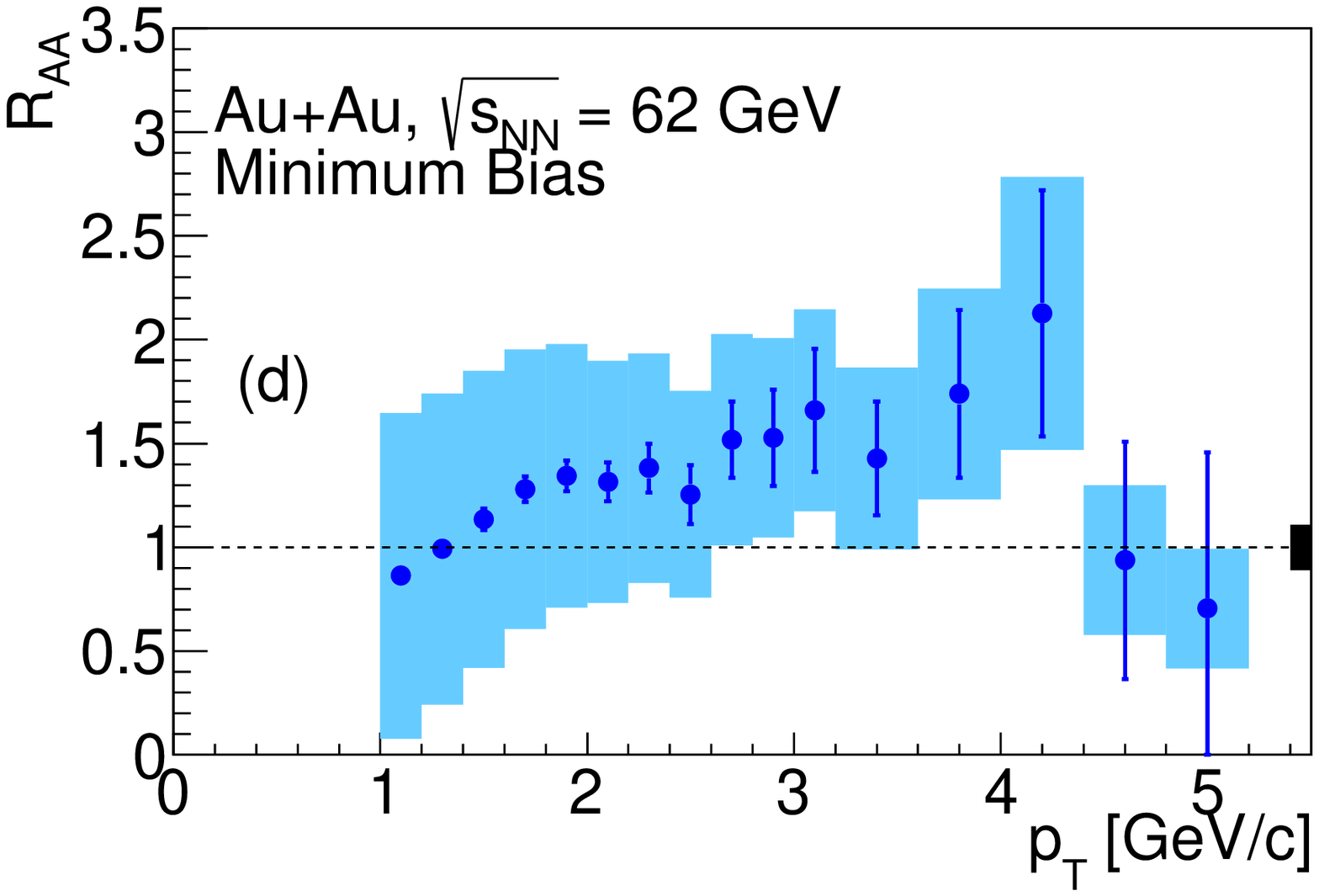}
\caption{\label{fig:RAA} (Color online) The \raa
for electrons from heavy-flavor decays in Au$+$Au collisions at 
\sqsnsix for the indicated centralities.  The error bars (boxes) represent the 
statistical (systematic) uncertainties.  The global uncertainty due to the 
uncertainty in $N_{\rm coll}$ for each centrality is given by the box on 
the right side of each plot.
}
\includegraphics[width=0.998\linewidth]{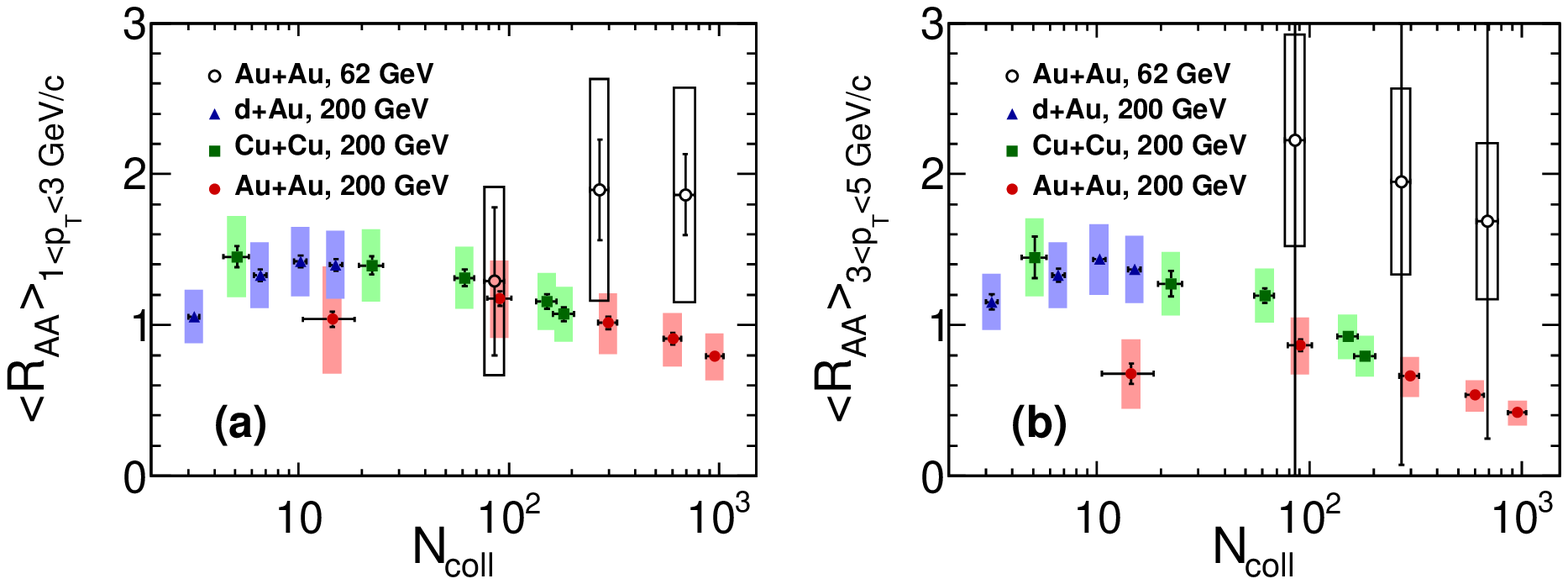}
\caption{\label{fig:RAA_all} (Color online)  
The \raa values for electrons from heavy-flavor decay in \auau 
collisions at 62.4~GeV with the \raa results from \dAu, Cu$+$Cu, 
and \auau collisions at \sqsntwo (data 
from~\cite{Adare:2010de,PhysRevLett.109.242301}). The error bars 
(boxes) represent the statistical (systematic) uncertainties. 
The \pt ranges are as indicated: (a) $1<p_T<3$~GeV/$c$ and (b) 
$3<p_T<5$~GeV/$c$.
}
 \end{figure*}



To compare the Au$+$Au data with $p$$+$$p$ results, we divide the Au$+$Au data 
by the number of binary collisions, \Ncoll. For each of the three 
centrality classes, Table \ref{tab:ncoll} lists the \Ncoll values. 
Figure~\ref{fig:yield_ncoll} compares the invariant yield of the 
heavy-flavor electrons per binary collision in 0\%--20\%, 20\%--40\%, 
40\%--60\% centrality bins and MB data in Au$+$Au collisions at \sqsnsix. 
The invariant cross section of heavy-flavor electrons in $p$$+$$p$ 
collisions at \sqsnsix is derived from the highest statistics 
heavy-flavor electron measurement~\cite{Basile1981} that was performed at 
the ISR. These results are scaled by the inelastic cross section at 
\sqsnsix, $\sigma_{pp}=35.9$~mb~\cite{Baksay19781}, and 
plotted in Fig.~\ref{fig:yield_ncoll}(e).

The 
fixed-order-plus-next-to-leading-log (FONLL) prediction~\cite{FONLL} 
(red curve) is also shown in Fig.~\ref{fig:yield_ncoll}. In Au$+$Au 
collisions at \sqsnsix, the yield of heavy-flavor electrons per binary 
collision is higher than the ISR results in $p$$+$$p$ collisions, while 
the ISR $p$$+$$p$ results are consistent with the upper limit of the FONLL prediction.

To further study the modification of the yield of heavy-flavor electrons 
in Au$+$Au collisions at \sqsnsix, the invariant yield per binary 
collision $N_{\rm coll}$ of heavy-flavor electrons is integrated across three 
\pt bins as shown in Fig.~\ref{fig:integrat_yield_only}. At $N_{\rm coll}=1$ 
the $p$$+$$p$ points come from the three published ISR 
measurements~\cite{Basile1981,Busser1976189,Perez1982260}.

At low \pt ($1.5<\pt<2.5$~GeV/$c$), an enhancement of the heavy flavor 
electron yield is observed in the 0\%--20\% and 20\%--40\% centrality bins 
relative to the yield in $p$$+$$p$ collisions, while the more peripheral 
40\%--60\% centrality bin is consistent with the $p$$+$$p$ yield, within 
uncertainties.  In the higher \pt ranges, $2.5<\pt<3.5$~GeV/$c$ and 
$3<\pt<5$~GeV/$c$, enhancement is observed relative to $p$$+$$p$ in all 
centrality bins.  A scenario with only heavy quark energy loss in a 
deconfined medium would show a pattern of increasing suppression with 
collision centrality, contrary to what is observed here. This suggests 
that other mechanisms are present.


We also calculate the nuclear-modification factor \raa, which is
the ratio of the yield per binary collision 
in \auau reactions divided by the yield in \pp collisions.  The \raa vs 
\pt are shown in Fig.~\ref{fig:RAA} for 3 different centrality classes and 
for MB.  The yield in \pp collisions is taken from the combined fit 
to the three ISR data sets~\cite{Basile1981,Busser1976189,Perez1982260}.  
The statistical uncertainty on \raa is taken from the statistical 
uncertainty on the heavy-flavor electron yield measured in Au$+$Au 
collisions shown in Fig \ref{fig:hfe_yield}.  The systematic uncertainty 
on \raa is a quadrature sum of the systematic uncertainty on the heavy 
flavor electron yield in Au$+$Au collisions and the statistical uncertainty on 
the fit used to represent the denominator.  At low \pt, where the fit to 
the $p$$+$$p$ denominator is relatively well constrained, the systematic 
uncertainty on \raa is dominated by the systematic uncertainty on the 
measured heavy-flavor electron yield in Au$+$Au.  At high \pt, where the 
$S/B$ ratio for heavy-flavor electrons in Au$+$Au collisions is relatively 
high and the fit representing the $p$$+$$p$ denominator is not well 
constrained, the systematic uncertainty on \raa is dominated by the 
uncertainty propagated from the fit parameters.  The \raa is consistently 
larger than unity with the exception of low-\pt data in peripheral collisions.   
In contrast to the heavy-flavor results, the $\pi^0$ data at 62~GeV show 
a suppression that increases with centrality~\cite{PhysRevLett.109.152301}.
 
\begin{figure}[htb]
\includegraphics[width=1.0\linewidth]{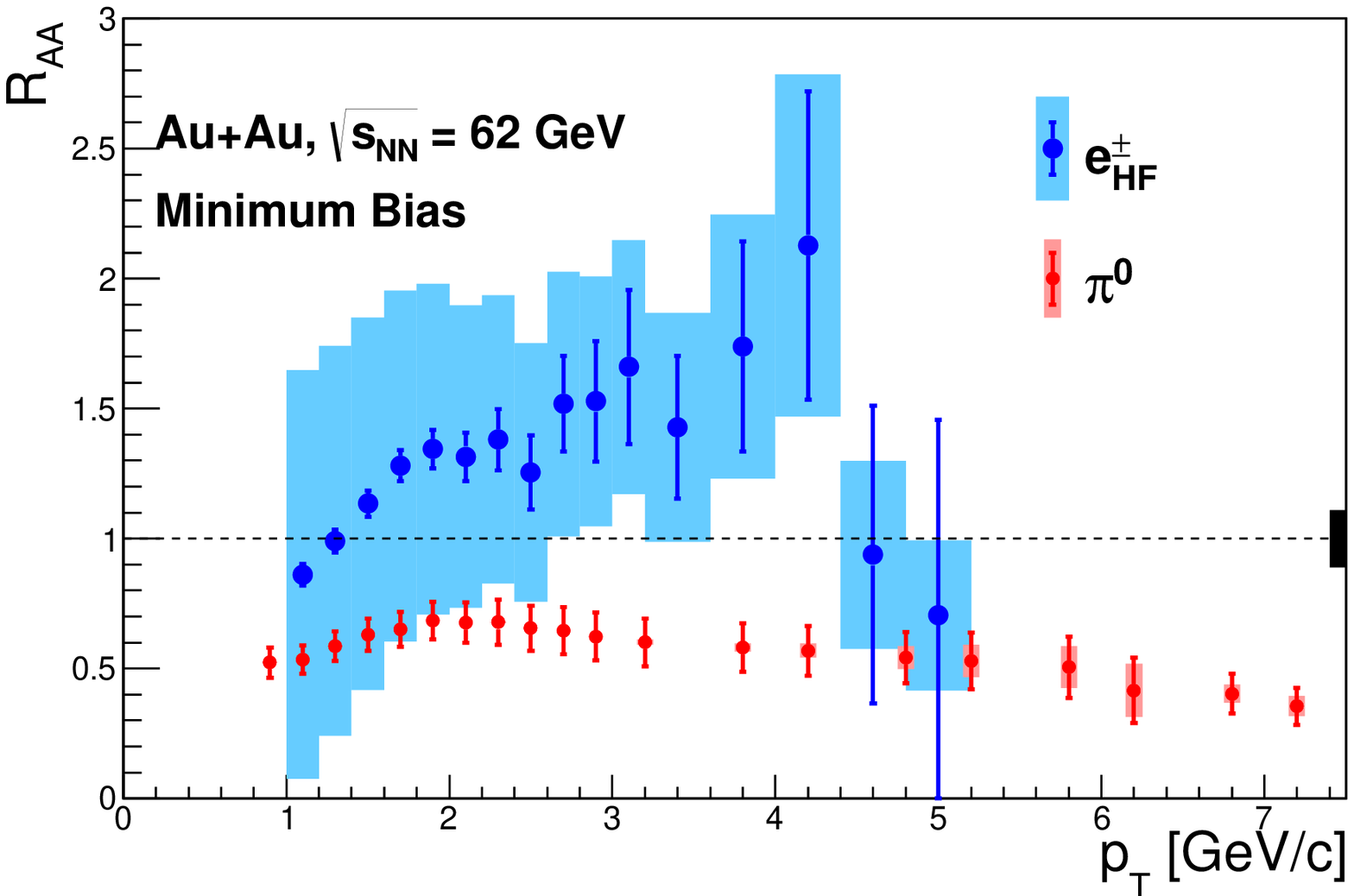}
\caption{\label{fig:RAA_HF_pi0}
(Color online)  The \raa values from heavy-flavor decay 
for electrons and $\pi^0$~\cite{PhysRevLett.109.152301} in \auau 
collisions at 62.4~GeV.  The error bars (boxes) show the statistical 
(systematic) uncertainties.}
 \end{figure}

These \raa values for electrons from heavy-flavor decay in \auau 
collisions at 62.4~GeV are compared to other \raa results from 
$d$$+$Au, Cu$+$Cu, and Au$+$Au collisions at \sqsntwo (data 
from~\cite{Adare:2010de,PhysRevLett.109.242301}), as shown in 
Fig.~\ref{fig:RAA_all}.  At 200~GeV the heavy-flavor \raa first 
increases with centrality then decreases, consistent with a competition 
between two mechanisms. At 62.4~GeV the competition, if present, favors 
heavy-flavor enhancement over suppression. This is consistent with 
previous results with hadrons where the Cronin enhancement increases as 
the collision energy decreases~\cite{Adler:2006xd}.
This competition between Cronin enhancement, flow, and suppression 
produces a different pattern for \raa for light mesons 
(Fig.~\ref{fig:RAA_HF_pi0}). 

\begin{figure}[htb]
\includegraphics[width=1.0\linewidth]{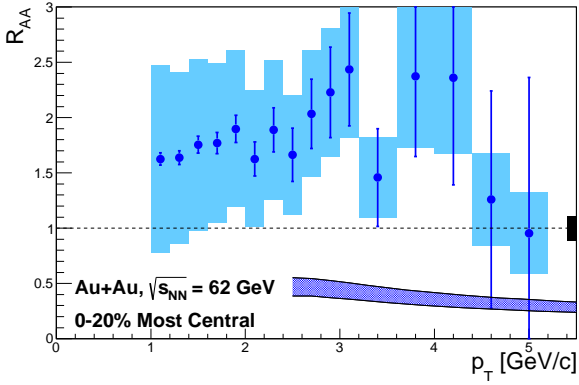}
\caption{\label{fig:RAA_model} (Color online)
The nuclear modification factor \raa for electrons from 
heavy-flavor decays in 0\%--20\% central \auau collisions at 
\sqsnsix compared to predictions (dark-blue band) from 
Vitev {\it et al.}~\protect\cite{Sharma:2009hn,Vitev2013}.
}
\end{figure}
\begin{figure}[htb]
 \includegraphics[width=1.0\linewidth]{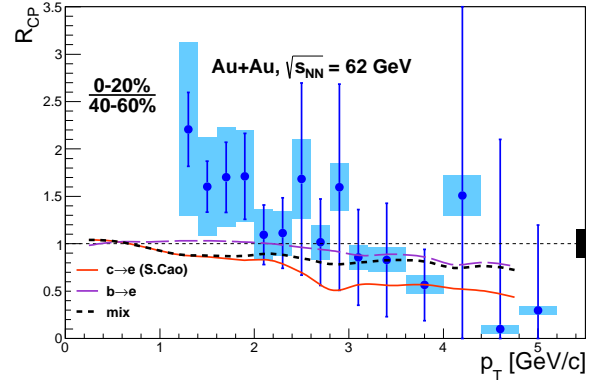}
 \caption{\label{fig:rcp} (Color online) 
Heavy-flavor electron \rcp between centrality 0\%--20\% and 
40\%--60\% in Au$+$Au collisions at \sqsnsix.
The curves are calculated using a
model based on energy loss~\protect\cite{Cao2013,Cao:2013ita}.
}
\end{figure}
\begin{figure}[htb]
\includegraphics[width=1.0\linewidth]{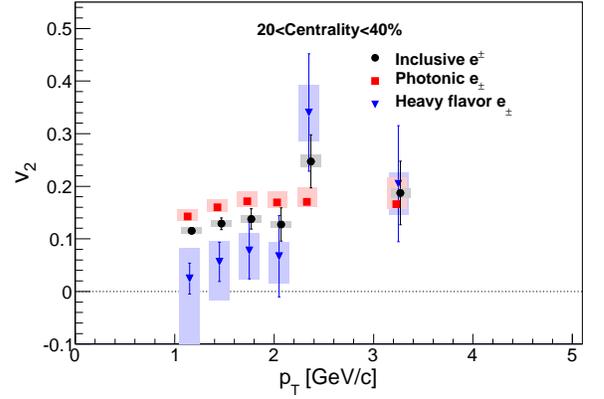}  
  \caption{\label{fig:inc_pho_v2}(Color online) 
Candidate (inclusive), photonic, and heavy-flavor electron 
$v_{2}$ in Au$+$Au collisions at \sqsnsix for 
20\%--40\% centrality.
}
\end{figure}
\begin{figure*}[htb]
\includegraphics[width=0.998\linewidth]{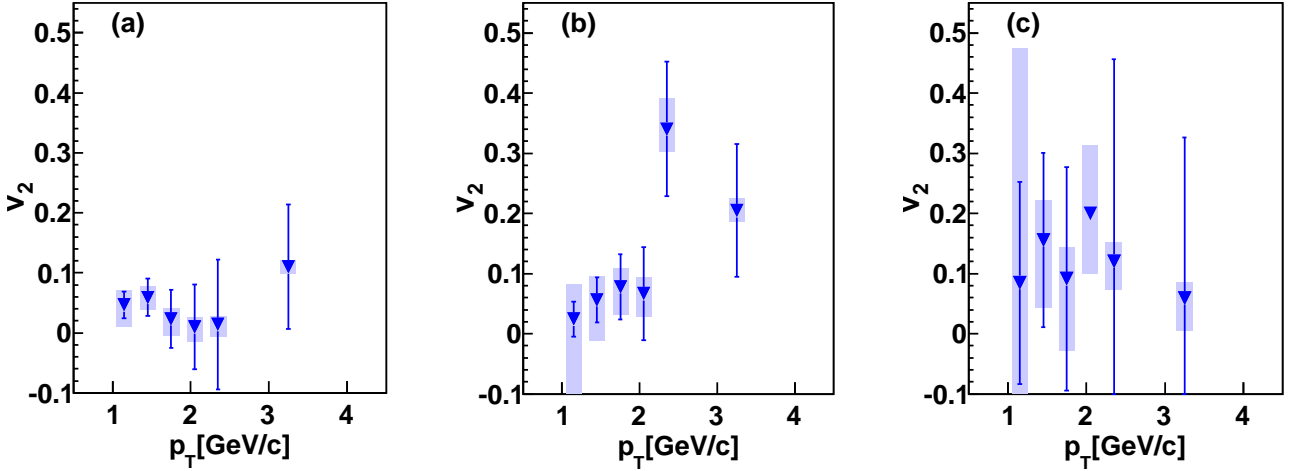}  
 \caption{\label{fig:hfe_v2}
(Color online) Heavy-flavor electron $v_{2}$ in Au$+$Au collisions at 
\sqsnsix for (a) 0\%--20\%, (b) 20\%--40\%, and (c) 40\%--60\% 
centrality bins.
}
 \end{figure*}
\begin{figure*}[htb]
\includegraphics[width=0.48\linewidth]{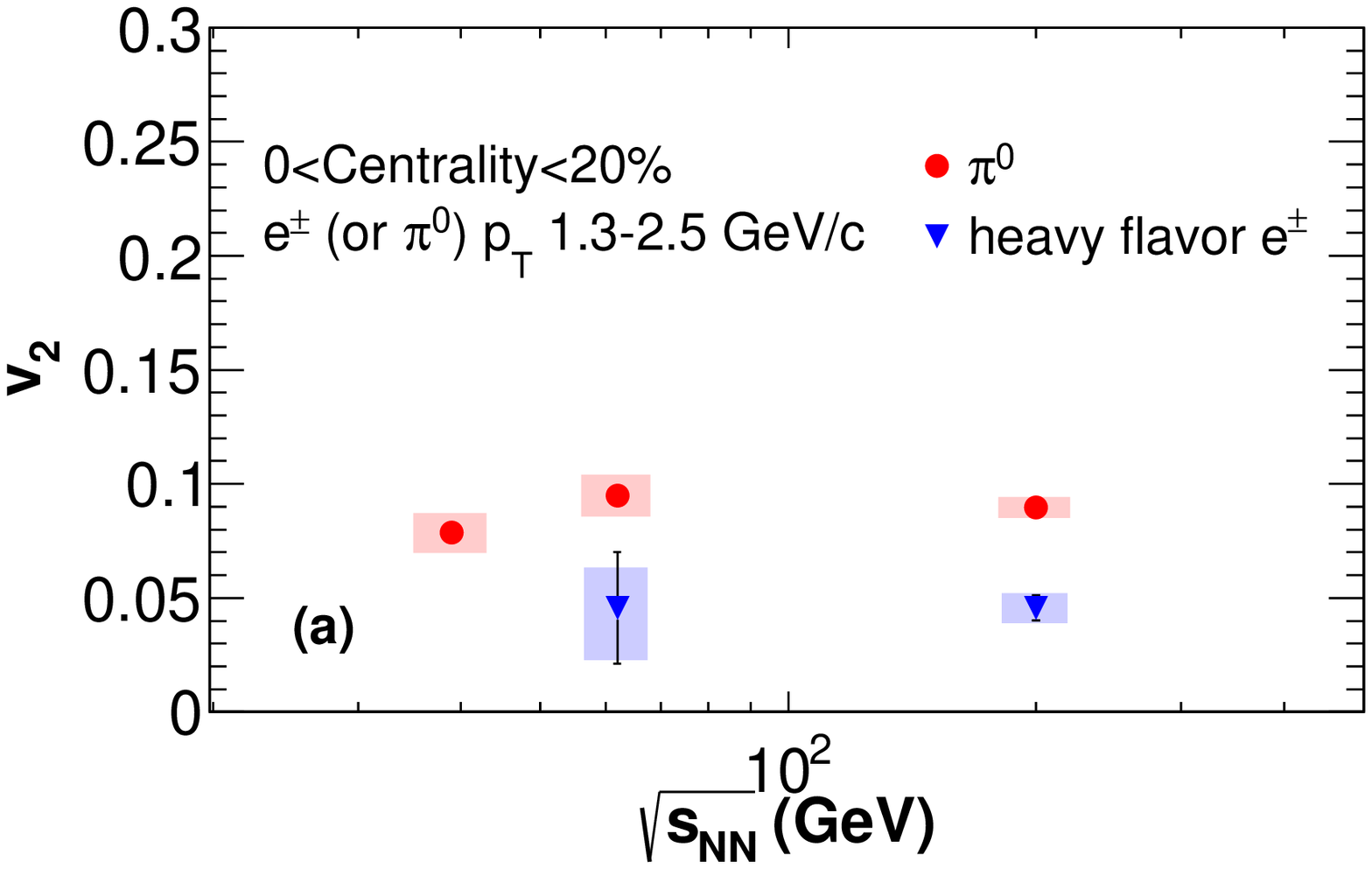}
\includegraphics[width=0.48\linewidth]{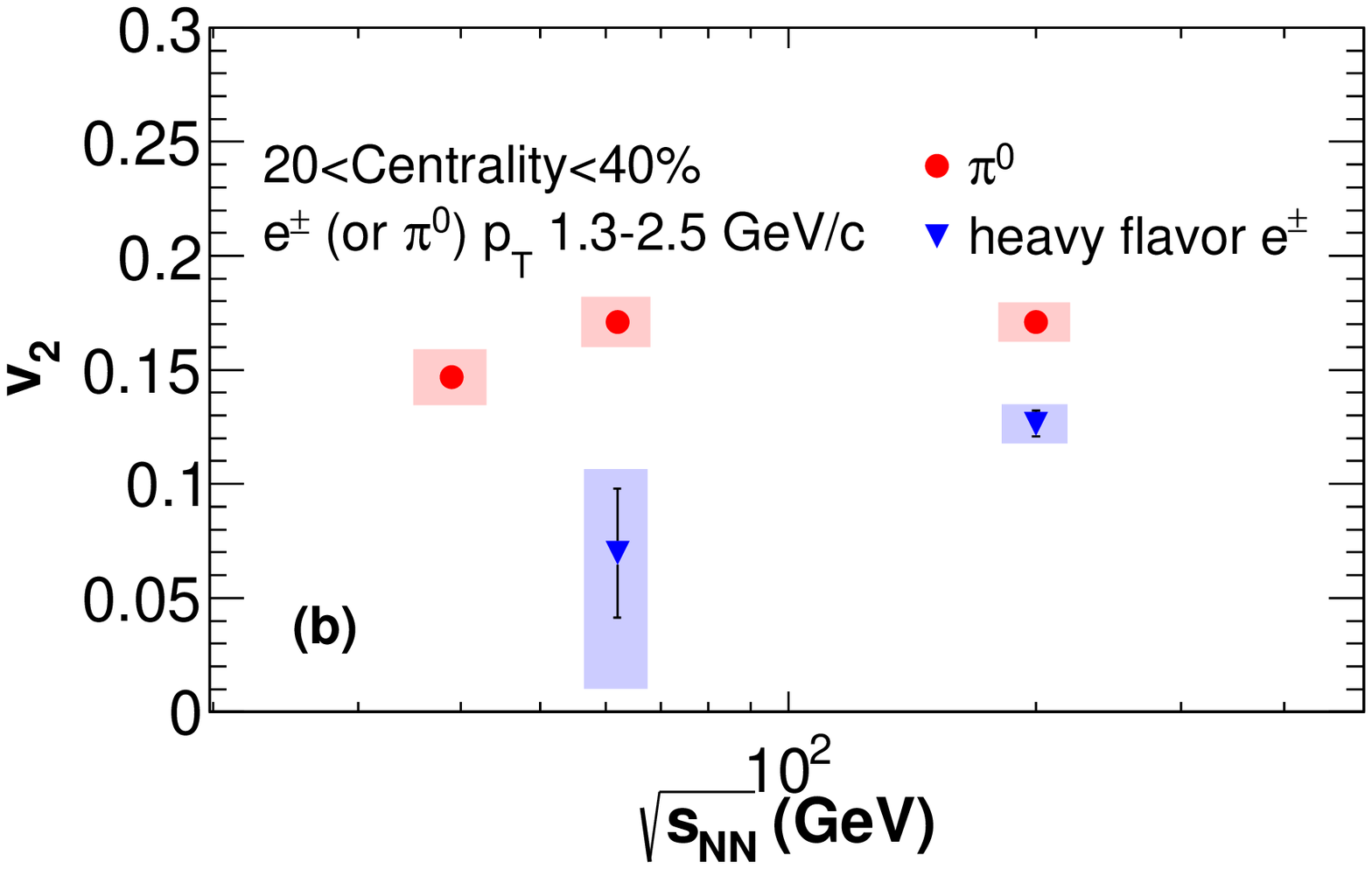}
\caption{\label{fig:excitation}
(Color online) The \vtwo of heavy-flavor electrons and $\pi^{0}$ 
in Au$+$Au collisions as a function of collision energy in 
the indicated \pt range of $1.3<p_T<2.5$~GeV/$c$ for 
(a) 0\%--20\% and (b) 20\%--40\% centrality.
}
 \end{figure*}

To estimate how rapidly the Cronin effect on heavy-flavor production could 
change from \sqsntwo to 62~GeV, we have performed {\sc pythia} calculations with 
different numerical $k_T$ parameters to estimate the possible size of the 
enhancement due to an increase in initial-state multiple scattering. 
Increasing $k_T$ from 0 to 1.5~GeV/$c$ enhances the yield of electrons from 
charm decay by a factor of 2.5 for $2<p_T<3$~GeV/$c$. At 200~GeV this 
enhancement is only a factor of 1.5. The observed enhancement of heavy 
flavor electrons could be due to less energy loss in the medium at 62.4 
GeV, a larger Cronin enhancement in the initial state at 62.4~GeV, or a 
combination of these factors. In addition to the Cronin enhancement, gluon 
anti-shadowing may increase the charm cross section in \auau collision at 
62.4~GeV, and cause the overall enhancement of the heavy-flavor electron 
yield per \Ncoll compared to scaled \pp collisions.  

Vitev has predicted \raa using his model of heavy-flavor 
energy-loss~\cite{Sharma:2009hn,Vitev2013}. Figure~\ref{fig:RAA_model} 
shows that these calculations, which include both energy-loss of 
heavy quarks inside a QGP as well as dissociation of D and B mesons, 
significantly underpredict the measured data.

As a complementary study of the change of the heavy-flavor electron yield
from peripheral to central collisions, we measure \rcp as defined by:
\begin{equation}
\rcp=\frac{\left<N_{\rm coll}^{\rm peripheral}\right> \times
dN_{\rm AuAu,central}^{e}/dp_{T}}{\left<N_{\rm coll}^{\rm central}\right> \times
dN_{\rm AuAu,peripheral}^{e}/dp_{T}}.
\label{eq:rcp}
\end{equation}
The yield from the 0\%--20\% centrality bin and 40\%--60\% centrality
bin are used for the numerator and denominator of \rcp respectively.
Fig.~\ref{fig:rcp} shows \rcp is above 1 for \pt below 1~GeV/$c$ and is 
consistent with 1 at higher transverse momenta.
The curves in Fig.~\ref{fig:rcp} are calculated using a
model based on energy loss~\cite{Cao2013,Cao:2013ita}.


\subsection{Heavy Flavor Electron $v_2$}

Heavy-flavor-electron \vtwo is calculated from 
candidate-electron \vtwo, photonic-electron \vtwo, and \rsb as:
\begin{equation}
v_{2}^{\rm hf}=v_{2}^{inc} (1+\frac{1}{\rsb})-v_{2}^{pho}\frac{1}{\rsb}.
\label{eq:hf_v2_eq}
\end{equation}


Figure~\ref{fig:inc_pho_v2} shows the measured \vtwo results for candidate 
electrons, photonic and heavy-flavor electrons in the 20\%--40\% 
centrality bin to illustrate their relative magnitude. 
Figure~\ref{fig:hfe_v2} shows the \vtwo of heavy-flavor electrons in 
Au$+$Au collisions at \sqsnsix in 0\%--20\%, 20\%--40\% and 40\%--60\% 
centrality bins. In the 20\%--40\% centrality bin, a nonzero \vtwo of 
heavy-flavor electrons is observed for $p_T>1.5$~GeV/$c$, which may 
indicate that charm quarks in the \pt range of this analysis experience 
some degree of collective motion along with the bulk medium.



To gain further insight into the possible differences in coupling to the 
medium due to quark mass, the \vtwo of heavy-flavor electrons and \pizero 
for $1.3<p_T<2.5$~GeV/$c$ in Au$+$Au collisions as a 
function of collision energy are compared in Fig.~\ref{fig:excitation}, 
for 0\%--20\% centrality and 20\%--40\% centrality.  The plots show that both 
heavy-flavor electrons and \pizero experience anisotropic flow in 
Au$+$Au collisions at \sqsnsix and 200~GeV.  The \vtwo for heavy-flavor 
electrons is lower than that for \pizero.  We note that the \pizero is a 
fully reconstructed meson, while the electrons from heavy-flavor decays 
are daughter products from the decay of charm and bottom mesons and 
baryons, and therefore the electron \pt does not necessarily represent the 
\pt of the parent hadron.

Because the heavy-flavor electrons are decay products from heavy flavor 
hadrons which may come from recombination of a heavy quark with a light 
quark from the bulk~\cite{Cao:2013ita}, heavy-flavor hadrons could acquire 
\vtwo as a consequence of recombination. Hence, a nonzero \vtwo of heavy 
flavor electrons does not necessarily imply a nonzero \vtwo of charm 
quarks. It will be necessary to compare our data with theoretical models 
with heavy quark flow for further understanding of the collective motion 
of the heavy quarks in the medium at 62.4 and 200~GeV.

\begin{figure}[htb]
\includegraphics[width=1.0\linewidth]{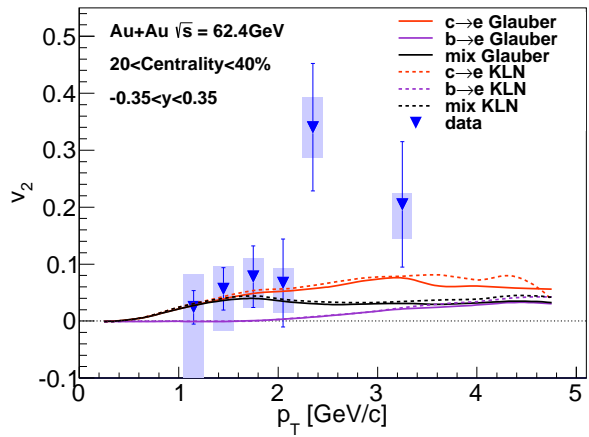}
 \caption{\label{fig:v2compshan}
(Color online) Heavy-flavor electron $v_{2}$ in Au$+$Au collisions at 
\sqsnsix compared with multiple theory 
curves~\protect\cite{Cao2013,Cao:2012au}.
}
\end{figure}

Figure~\ref{fig:v2compshan} shows such a comparison between our \vtwo 
results and theoretical calculations~\cite{Cao2013,Cao:2013ita}, which
use the framework of a modified Langevin equation~\cite{Cao:2012au} coupled 
to a (2+1)-dimensional viscous hydrodynamic model~\cite{Shen:2012vn}. The 
classical Langevin approach is improved by adding both quasi-elastic 
scattering and medium-induced gluon radiation for heavy quark energy loss 
inside the QGP medium. Before the Langevin evolution, heavy quarks are 
initialized with a leading order perturbative quantum chromodynamics 
calculation~\cite{RevModPhys.59.465} coupled to the nuclear parton 
distribution function provided in~\cite{Eskola:2008ca}. After traversing the 
QGP, the heavy quarks hadronize into heavy mesons according to a hybrid 
model of instantaneous coalescence~\cite{Oh:2009zj} and {\sc 
pythia}~6.4~\cite{Sjostrand:2006za} fragmentation. One set of initial 
conditions for the hydrodynamic model is used here, 
MC-Glauber~\cite{Miller:2007ri}.  The calculations are in good agreement 
with the experimental data up to $p_{T}=2$~GeV/$c$.

Two initial conditions for the hydrodynamic model, 
MC-Glauber~\cite{Miller:2007ri} and KLN-CGC~\cite{Kharzeev:2004if} are 
compared in Fig.~\ref{fig:v2compshan} and the corresponding impact on the 
final state heavy flavor spectra is displayed. The \vtwo predictions in 
the model show nonzero flow for electrons from heavy-flavor hadrons 
(Fig.~\ref{fig:v2compshan}), which are mainly D mesons for 
\pt$<$5~GeV/$c$.  This model is consistent with the \vtwo data at low \pt, 
within experimental uncertainties.

\section{Summary and Conclusions}
\label{sec04}

This article presents the measurements of the invariant yield and elliptic 
flow of electrons from heavy flavor meson semi-leptonic decays in \auau 
collisions at \sqsnsix in PHENIX. The integrated invariant yield per 
binary collision is slightly larger than the yields from prior \pp 
measurements. This enhancement is different from the suppression observed 
in previous PHENIX measurements of heavy-flavor electrons in \auau at 
\sqsntwo, but is comparable to the enhancement observed in \dAu collisions 
at \sqsntwo.  Hence it is possible that the initial state Cronin 
enhancement becomes the dominant effect at low to moderate \pt for heavy 
quarks at this lower beam energy compared to energy loss in the medium. 
The measured \vtwo of heavy-flavor electrons is positive when averaged 
across $p_T$ between 1.3 and 2.5~GeV/$c$.  The heavy-flavor \vtwo is 
smaller than the $\pi^0$ \vtwo, and may be caused by collective motion of 
charm quarks themselves and/or charmed hadrons accruing collective motion 
through recombination with flowing light partons. Further understanding of 
the properties of the medium and energy loss of the heavy quarks at 62.4 
GeV requires the measurement of cold nuclear matter effects on heavy 
flavor through \pp or \dA collisions at 62.4~GeV, as well as a separation 
of the individual contributions from charm and bottom hadrons.


\section*{ACKNOWLEDGMENTS}


We thank the staff of the Collider-Accelerator and Physics
Departments at Brookhaven National Laboratory and the staff of
the other PHENIX participating institutions for their vital
contributions.  We acknowledge support from the 
Office of Nuclear Physics in the
Office of Science of the Department of Energy, the
National Science Foundation, Abilene Christian University
Research Council, Research Foundation of SUNY, and Dean of the
College of Arts and Sciences, Vanderbilt University (U.S.A),
Ministry of Education, Culture, Sports, Science, and Technology
and the Japan Society for the Promotion of Science (Japan),
Conselho Nacional de Desenvolvimento Cient\'{\i}fico e
Tecnol{\'o}gico and Funda\c c{\~a}o de Amparo {\`a} Pesquisa do
Estado de S{\~a}o Paulo (Brazil),
Natural Science Foundation of China (P.~R.~China),
Croatian Science Foundation and 
Ministry of Science, Education, and Sports (Croatia),
Ministry of Education, Youth and Sports (Czech Republic),
Centre National de la Recherche Scientifique, Commissariat
{\`a} l'{\'E}nergie Atomique, and Institut National de Physique
Nucl{\'e}aire et de Physique des Particules (France),
Bundesministerium f\"ur Bildung und Forschung, Deutscher
Akademischer Austausch Dienst, and Alexander von Humboldt Stiftung (Germany),
Hungarian National Science Fund, OTKA (Hungary), 
Department of Atomic Energy and Department of Science and Technology (India), 
Israel Science Foundation (Israel), 
National Research Foundation of Korea of the Ministry of Science,
ICT, and Future Planning (Korea),
Physics Department, Lahore University of Management Sciences (Pakistan),
Ministry of Education and Science, Russian Academy of Sciences,
Federal Agency of Atomic Energy (Russia),
VR and Wallenberg Foundation (Sweden), 
the U.S. Civilian Research and Development Foundation for the
Independent States of the Former Soviet Union, 
the Hungarian American Enterprise Scholarship Fund,
and the US-Israel Binational Science Foundation.




%

\end{document}